\documentclass[pre,preprintnumbers,amsmath,amssymb,superscriptaddress]{revtex4}
\usepackage{amsmath}
\usepackage{mathtools}
\usepackage{rotating}
\usepackage{pdflscape}
\usepackage{array}
\usepackage{graphicx}
\usepackage{stmaryrd}
\usepackage{dcolumn}
\usepackage{booktabs}
\usepackage{bm}
\usepackage{longtable}

\DeclareRobustCommand{\stirling}{\genfrac\{\}{0pt}{}}

\usepackage{hyperref}
\usepackage{multirow}
\hypersetup{
	colorlinks=true,
	linkcolor=blue,
	citecolor=blue,
	urlcolor=blue
}
\usepackage{epstopdf}
\allowdisplaybreaks

\begin{document}

\preprint{APS/123-QED}

\title{Beyond the Big Jump: A Perturbative Approach to Stretched-Exponential Processes}

\author{Alberto Bassanoni}
\thanks{Corresponding author: \href{mailto:alberto.bassanoni@unipr.it}{alberto.bassanoni@unipr.it}}
\affiliation{Dipartimento di Scienze Matematiche, Fisiche ed Informatiche,
Università degli Studi di Parma, Parco Area delle Scienze 7/A, 43124, 
Parma, Italy}
\affiliation{INFN, Gruppo Collegato di Parma, Università degli Studi di Parma, Parco Area delle Scienze 7/A, 43124, Parma, Italy}

\author{Omer Hamdi}
\affiliation{Department of Physics, Institute of Nanotechnology and Advanced Materials, Bar-Ilan University, Ramat-Gan, 52900, Israel}

\begin{abstract}
The problem of sums of independent, identically distributed random variables with stretched-exponential tails exhibits a dynamical phase transition and has recently reemerged in the context of active transport and condensation phenomena. We develop a perturbative expansion for the distribution of the sum that systematically extends the Big Jump Principle beyond its asymptotic regime. The expansion yields explicit higher order corrections that describe moderate deviations, bridging the gap between typical Gaussian fluctuations and the far-tail behavior dominated by single big jump events. In this sense, our approach is complementary to the classical Edgeworth expansion, which provides corrections to the Gaussian core, whereas we construct systematic corrections to the big jump regime. The leading terms reveal the scaling structure governing the crossover between typical and condensed fluctuations, in agreement with large deviation predictions but without relying on its asymptotic limit. We further extend the framework to continuous-time random walks (CTRWs), where stretched-exponential jump statistics combined with stochastic renewal times generate nontrivial propagators through subordination. This setting is particularly relevant for transport processes with non-Gaussian displacement statistics, where super-exponential or Laplace-like tails emerge from the interplay of rare large jumps and temporal fluctuations. All analytical predictions are supported by numerical simulations.
\end{abstract}

\maketitle

\section{Introduction}

The study of rare events of a sum of independent and identically distributed random variables (IID RVs) is a well-known problem that mathematicians have widely studied since the early 20th century. In recent years, studies of the probability density function (PDF) of a sum of heavy-tailed IID RVs, particularly the sum of RVs extracted from stretched-exponential distributions, have become very relevant. Some recent rigorous mathematical works based on the theory of large deviations \cite{LDT_00, LDT_0, LDT_1, LDT_2, LDT_3, LDT_4, LDT_5, LDT_6} discovered interesting features on the asymptotic behavior of these distributions. In the limit of a large number of IID RVs, the PDF of their sum is characterized by two different asymptotic regimes: below a certain typical scale length of the system a Gaussian regime is found and the PDF of the sum follows the usual prescription of the Central Limit Theorem (CLT); conversely, beyond this typical length, the PDF of the sum is governed by the single Big Jump Principle (BJP) \cite{BJ_theorem, BJ_Foss, BJ_1, BJ_2, BJ_3, BJ_4, BJ_5, BJ_6, BJ_subexp, BJ_subexp_2, BJ_FPT}. The two asymptotic regimes are separated by a first-order phase transition in the limit of an infinite number of RVs, and it is possible to write, through large deviation arguments, a rate function describing the logarithm of the full asymptotic PDF of the sum and its transition, rescaling it with respect to the typical length scale of the system.

These processes have long been studied in the physical sciences, as they are excellent prototypes for describing transport processes subject to condensation phenomena. In a generic many-body system, a condensation transition occurs when, studying an additive observable of the system dependent on all elements' contributions, a single one captures a significant fraction of the value of the observable, i.e., beyond a certain critical value, the system’s contribution “condensates” into a single element’s contribution. There is a large variety of examples in different microscopic models, such as run-and-tumble particle systems \cite{Condensation_run_and_tumble_1, Condensation_run_and_tumble_2, Condensation_run_and_tumble_3, Condensation_run_and_tumble_4, Condensation_run_and_tumble_5}, active matter \cite{Active_matter_1, Active_matter_2, Active_matter_3, Active_matter_4, Active_matter_5}, random matrix theory \cite{Random_matrix_1, Random_matrix_2, Random_matrix_3}, and overdamped Langevin processes \cite{overdamped_OU_1, overdamped_OU_2, overdamped_OU_3, overdamped_OU_4, overdamped_OU_5, overdamped_OU_6, overdamped_OU_7}, in which one can also include the effects of stochastic resetting \cite{Stochastic_resetting_1, Stochastic_resetting_2, Stochastic_resetting_3, Stochastic_resetting_4}.

The tail events in the condensed phase have already been studied in the context of CTRWs \cite{BJ_subexp, Omer_ref}, and they are governed by the BJP \cite{BJ_theorem}, which states that the probability of extracting a certain value for the sum is asymptotically equal in the tail regime to the probability of having a single RV of that value, i.e., it is the probability of making a single big jump multiplied by the number of attempts in which it can occur, which coincides with the number of RVs.

However, this result holds only in the far-tailed regime, and it is not clear what happens in the region near the phase transition, where our process is still outside the Gaussian attractor of the CLT but at the same time is far from the asymptotic BJP regime. 
Systematic corrections to the central region of the distribution can be described by Edgeworth-type expansions. For heavy-tailed or non-Gaussian attractors, these ideas have been generalized to fractional Edgeworth expansions, which provide controlled corrections to Gaussian-L\'evy CLT behavior and capture finite-size deviations in the bulk of the distribution \cite{Edgeworth_1, Edgeworth_2}. Our work fits within this framework and aims to capture the behavior of the PDF of the sum of stretched-exponential IID RVs in this intermediate regime, making use of a perturbative expansion around the BJP solution, extending the previous results obtained in \cite{Omer_ref}, and providing an analytical expression for each perturbative order. In this sense, our approach plays a role complementary to Edgeworth-type expansions: while Edgeworth theory provides systematic corrections around the typical fluctuations described by the CLT, our expansion provides systematic corrections around the rare-event configurations underlying the BJP regime.

Recently, a perturbative expansion for the sum of IID RVs was proposed in Ref.~\cite{LDT_Perturbative_Naftali}, where the author investigated the same physical problem from the complementary viewpoint of the large-deviation formalism. In his work, the expansion is performed in powers of the number of summed RVs $1/n$, starting from the CLT regime in the limit of large $n$ and building a sub-leading theory that connects it to the BJP phase. In contrast, our perturbative expansion is developed in powers of $1/x$, in the limit of large $x$, directly around the BJP solution. The two approaches are therefore complementary: while the first theory provides a large $n$ expansion from the Gaussian side, ours builds a large $x$ expansion from the condensation side.

As in the case of the classical Edgeworth expansion, such perturbative descriptions remain valuable even when exact large-deviation principles are known, since they provide explicit and accurate approximations in regimes where asymptotic theories may converge slowly or apply only in limiting cases. Importantly, from the leading terms of our large $x$ expansion we recover a scaling relation that represents the approximate rate function near the dynamical phase transition between the condensed phase and the typical-fluctuation phase, in consistency with the large-deviation results obtained in \cite{overdamped_OU_4, LDT_Perturbative_Naftali}. Moreover, our perturbative framework can be naturally extended to CTRWs with stretched-exponential jump-length PDFs, for which we derive the analytical structure of the propagator’s PDF. All our analytical results will be tested with extensive numerical simulations.

The paper is organized as follows. In Section \ref{Section2}, we introduce the model and present the main results for sums of IID RVs. In Section \ref{Section3}, we detail the derivation of the perturbative series around the BJP. In Section \ref{Section4}, we analyze the leading contributions of the series and their connection with the anomalous rate function describing the dynamical phase transition between the CLT and BJP regimes. Finally, in Section \ref{Section5}, we extend our approach to decoupled CTRWs and discuss its implications for systems with stretched-exponential transport.

\section{Model and Main Results}
\label{Section2}

We consider a one–dimensional CTRW in which the motion consists of a sequence of instantaneous spatial jumps separated by random waiting times. The waiting times 
$\{\tau_i\}_{i=1}^n$
are IID RVs drawn from a waiting time PDF \(\psi(\tau)\), which we assume to have a finite mean waiting time \(\langle \tau \rangle < \infty\). The jump lengths $\{\chi_i \}_{i=1}^n$
are IID RVs drawn from a symmetric stretched-exponential distribution, defined as:
\begin{equation}
\label{f_x}
    f(\chi) = \widetilde N \exp\!\left( - \alpha^\beta |\chi|^\beta \right), \qquad 0<\beta<1.
\end{equation}
The parameter $\alpha$ and the normalization $\tilde{N}$ are chosen so that \(f(\chi)\) has zero mean and unit variance, namely $\alpha = \sqrt{\frac{\Gamma(3/\beta)}{\Gamma(1/\beta)}}$ and $\tilde{N} = \frac{\beta}{2} \frac{\sqrt{\Gamma(3/\beta)}}{\Gamma(1/\beta)^{3/2}}$, where \(\Gamma(\cdot)\) denotes the Gamma function. After exactly \(n\) jumps, the position of the walker is $x = \sum_{i=1}^n \chi_i$,
while the operational time is $t = \sum_{i=1}^n \tau_i$. For a given observation time $t$, the number of jumps performed up to time $t$ is the random variable $n_t = \max{n : t_n \le t}$, so that in general $t_n \le t < t_{n+1}$. The time elapsed since the last jump (the backward recurrence time) is therefore not included in $t_n$, but is implicitly accounted for through the counting statistics of $n_t$. Since the waiting-time statistics determine the number of jumps performed up to time \(t\), CTRWs are semi-Markov processes naturally treated within renewal theory \cite{Renewal_1, Renewal_2, Renewal_3}. Let \(Q_t(n)\) denote the probability of performing exactly \(n\) jumps up to time \(t\), , i.e.
$Q_t(n) = \text{Prob}(n_t = n)$.
The distribution $Q_t(n)$ is fully determined by the waiting-time PDF $\psi(\tau)$ through renewal theory; in particular, its moments describe the statistics of the random number of renewal events in the time interval $[0,t]$. Conditioning on \(n\), the propagator of the CTRW, i.e. the PDF to be at position $x$ at time $t$ $P(x,t)$, satisfies the standard subordination identity
\begin{equation}
\label{subordination}
    P(x,t) = \sum_{n=0}^{\infty} Q_t(n)\, \phi(x|n),
\end{equation}
where the conditional PDF to perform a total displacement $x$ after $n$ jumps $\phi(x|n) = f^{*n}(x)$
is the \(n\)-fold convolution of the jump PDF. By symmetry of \(f(\chi)\), both \(\phi(x|n)\) and \(P(x,t)\) are even functions of \(x\). The purpose of this section is twofold. First, we summarize the two limiting behaviors of \(\phi(x|n)\) and \(P(x,t)\), namely the Gaussian regime described by the CLT and the far-tail regime governed by the BJP. Second, we explain why a perturbative approach is required to interpolate between these regimes and introduce the geometric framework underlying our expansion.

\subsection{Two limiting behaviors: CLT vs BJP}

For large numbers of jumps $n \gg 1$, every sum of IID RVs with finite variance satisfies the CLT. Thus, for $(x| \ll \sqrt{n}$ the position of the walker $x$ after $n$ jumps is:
\begin{equation}
\label{phi_x_n_clt}
\phi(x|n)_{CLT} \approx \frac{1}{\sqrt{2\pi n}}
\exp\left(-\frac{x^2}{2n}\right).
\end{equation}
If the mean waiting time of the jumps $\langle \tau \rangle$ is finite, then the typical number of jumps performed by time $t$ is
$\langle n_t \rangle = \sum_{n=0}^\infty n, Q_t(n)$,
and for long times one has the renewal-theory estimate
$\langle n_t \rangle \simeq t/\langle \tau \rangle$.
Accordingly, the propagator of the CTRW $P(x,t)$ reduces to the Gaussian form
\begin{equation}
P(x,t)_{CLT} \approx \frac{1}{\sqrt{4\pi D t}}
\exp\left( -\frac{x^2}{4Dt} \right),
\qquad D = \frac{1}{2\langle \tau \rangle}.
\end{equation}
This regime describes the typical fluctuations of both $\phi(x|n)$ and $P(x,t)$.
For stretched-exponential jump distributions with $0<\beta<1$, large displacements in the tail regime are instead produced by a single exceptionally large jump. This is the content of the BJP, according to which
\begin{equation}
\label{phi_x_n_BJP}
\phi(x|n)_{BJP} \sim n f(x), \qquad |x|\to \infty,
\end{equation}
for any finite value of $n$. This asymptotic form reflects the fact that extreme displacements are dominated by a single jump whose size is of order $x$, while the remaining $n-1$ jumps contribute only subleading corrections.
Inserting this into the subordination relation yields:
\begin{equation}
P(x,t)_{BJP} \sim \langle n_t \rangle f(x), \qquad |x|\to\infty.
\end{equation}
This result expresses the fact that, in a CTRW where the number of jumps is itself random, the tail of the propagator is governed by the expected number of opportunities for a single large jump to occur. Similar formulations of the BJP for random numbers of summands appear in the renewal-process and applied-probability literature \cite{BJP_CTRW_0, BJP_CTRW_1, BJP_CTRW_2}. 

The CLT and BJP describe opposite extremes of the distribution. Between them lies a broad intermediate regime associated with a condensation transition \cite{LDT_5}, where the behavior is neither Gaussian nor fully dominated by single big jump events. In this regime, the total displacement typically results from the coexistence of one large jump and many smaller ones, and neither limiting asymptotic description is sufficient. Understanding this crossover for a finite number of jumps $n$ therefore requires a perturbative theory.

\subsection{Introducing perturbations}

\begin{figure}[t]
    \centering
    \includegraphics[width=0.9\textwidth]{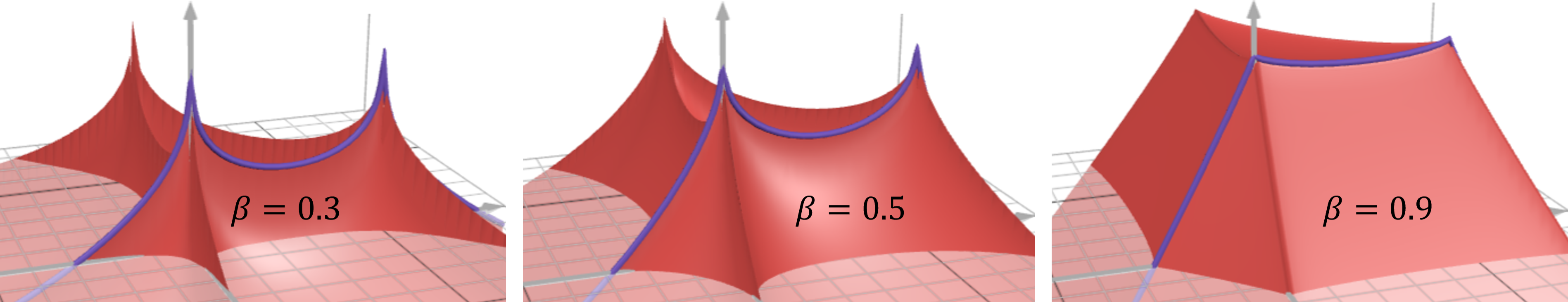}
    \caption{
    Graphical representation of \(g(\vec{y})\) for \(n=3\) jumps (red surface), with an overlay of \(g(\vec{y})\) for \(n=2\) (blue line), shown for increasing values of \(\beta = \{0.3,\,0.5,\,0.9\}\).
    The function exhibits one central cusp at \(\vec y = 0\) and \(n-1\) outskirt cusps located along the integration axes, each corresponding to a distinct realization of the Big Jump event occurring in one of the summands.  
    As \(\beta \to 1^{-}\), the stretched-exponential distribution approaches a piecewise-linear form, and the cusps become more linear; the convergence to the leading BJP term \eqref{phi_x_n_BJP} correspondingly slows down.  
    }
    \label{fig:cusps_3d}
\end{figure}

The CLT and the BJP are both correct asymptotically, but neither provides an adequate description of the intermediate region where stretched-exponential statistics crosses over from Gaussian behavior to big jump dominance. Our goal is to construct a systematic expansion around the big jump configuration that captures finite $x$ corrections generated by the remaining $n-1$ summands. Convergence to the BJP is particularly slow as $\beta \to 1^{-}$, making a systematic expansion necessary. This intermediate regime is also the one described by large deviation theory in the limit $n\to\infty$ \cite{LDT_5, overdamped_OU_4, LDT_Perturbative_Naftali}; here we instead develop an expansion valid for large $x$ at finite $n$. Our starting point is the convolutional interpretation of the PDF $\phi(x|n)$. Writing the convolution explicitly and rescaling the $n-1$ random jumps via $y_i=\chi_i/x$, one obtains the exact representation
\begin{equation}
\label{convolution}
\phi(x|n)
= |x|^{\,n-1}\tilde N^n
\int dy_1 \cdots \int dy_{n-1}\,
\exp\!\left[-\alpha^\beta |x|^\beta g(\vec y)\right],
\end{equation}
which involves $n-1$ integration variables corresponding to the $n-1$ summands that do not realize the big jump, where
\begin{equation}
\label{g_y_n}
g(\vec y)
= \sum_{i=1}^{n-1} |y_i|^\beta
+ \left|1-\sum_{i=1}^{n-1} y_i\right|^\beta.
\end{equation}
Figure \ref{fig:cusps_3d} illustrates the shape of $g(\vec{y})$ for increasing values of \(\beta<1\) for \(n=3\) jumps, and it is characterized by several cusps. The function $g(\vec y)$ has exactly $n$ non-analytic cusps. Physically, each cusp represents one of the $n$ distinct ways in which a single displacement of order $x$ can dominate the entire sum, i.e. the $n$ possible equivalent ways to perform a single big jump (in Appendix \ref{AppendixA} one can find a brief discussion about the symmetries of the $n$ cusps). In the large $x$ limit, the integral \eqref{convolution} is dominated by contributions from small neighborhoods of these cusps. The dominant contribution comes from one jump of order $x$, while the remaining $n-1$ jumps generate subleading corrections through their local fluctuations around the cusp.
\\
Expanding $g(\vec y)$ in the vicinity of each cusp yields a systematic expansion in powers of $1/x^2$, which originates from the large pre-factor $|x|^\beta$ in the exponential of Eq. \eqref{convolution}, and the full derivation of this expansion for the case of $n=2$ and its generalization to arbitrary $n$ is reported in Section \ref{Section3}. The zeroth order of this expansion reproduces the BJP, while higher orders describe correlations related to the cumulative effect of the $n-1$ remaining summands surrounding the big jump. The resulting large $|x|$ approximation takes the compact form
\begin{eqnarray}\label{phi_x_n_pert}
\phi(x|n)
\sim n f(x)\,[G(x)]^{\,n-1} , \qquad
G(x)
= 1+\sum_{\text{$l>0$, $l$ even}}^\infty
\frac{d_l(x)}{x^l}\,\mathbb{E}[x^l],
\qquad 
\mathbb{E}[x^l] = 
\begin{cases}
0 & \text{if $l$ odd} \\
\frac{\Gamma\left(1/\beta\right)^{l/2 - 1}}{\Gamma(3/\beta)^{l/2}}\Gamma\left( \frac{l+1}{2} \right) & \text{if $l$ even}
\end{cases}.
\end{eqnarray}

The function $G(x)$ does not depend on $n$; it depends only on the large $x$ expansion coefficients $d_l(x)$ and on the moments of the jump distribution. The coefficients $d_l(x)$ arise from the local expansion of $g(\vec y)$ around a cusp and depend only on $x$ and $\beta$. Their explicit expression is
\begin{equation}
\label{d_l_x_general_main}
d_l(x)
= \sum_{j=1}^l
\sum_{\substack{j_1+\cdots+j_l=j\\
j_1+2j_2+\cdots+lj_l=l}}
\frac{(-1)^{j+l}}{j_1!\cdots j_l!}
\,\alpha^{j\beta}|x|^{j\beta}
\binom{\beta}{1}^{j_1}
\binom{\beta}{2}^{j_2}\cdots
\binom{\beta}{l}^{j_l},
\end{equation}
The full derivation is provided in Appendix \ref{AppendixB}. The quantities $\mathbb{E}[x^l]$ denote the $l$-th moments of the PDF \eqref{f_x}, which are directly related to the cumulants appearing in the classical Edgeworth expansion. Since $f(x)$ is an even function, the perturbative series is non-trivial only for even $l>0$, as the moments of $x$ are null for every odd value of $l$.
The series in Eq. \eqref{phi_x_n_pert} should be viewed as an asymptotic expansion in the large $x$ limit. As in the case of the classical Edgeworth expansion, we use it in practice by truncating the series at finite order, and we do not discuss the question of convergence of the infinite series from a mathematical point of view. However, in order to perform our numerical simulations efficiently, we build an optimal truncation method that selects the appropriate perturbative order $L$ as $x$ varies, thus making $L=L(x)$ and providing an efficient computational recipe, maintaining the expansion in its radius of convergence, and this is discussed in Subsection \ref{Subsection3C}. 

We will show in Section \ref{Section4} that the leading terms of the large $x$ expansion \eqref{phi_x_n_pert} satisfy a scaling relation that contains the anomalous rate function describing the phase transition between the CLT and BJP regimes. This connects our large $x$ perturbative framework to the complementary large $n$ perturbative approach of Ref.~\cite{LDT_Perturbative_Naftali}.

Finally, since our formalism is built from the conditional distribution $\phi(x|n)$, it naturally extends to CTRWs with arbitrary waiting time PDFs, yielding analytical predictions for the propagator $P(x,t)$ also in the moderate-deviation regime. The derivation of the CTRW result below follows from inserting Eq.~\eqref{phi_x_n_pert} into the subordination relation \eqref{subordination} and resumming the series over $n$; the full derivation is presented in Section \ref{Section5}. This yields the perturbative expansion of the CTRW propagator in the large $x$ limit,
\begin{equation}
\label{P_x_t_pert}
P(x,t)\sim
\langle n_t\rangle f(x)
+\sum_{m=1}^\infty \frac{1}{m!}
\left[\sum_{k=0}^m(-1)^k{m\choose k}
\langle n_t^{\,m-k+1}\rangle\right]
[G(x)-1]^m f(x),
\qquad |x|\to\infty.
\end{equation}
This result applies to any CTRW with finite mean waiting time $\langle\tau\rangle$, regardless of the detailed form of $\psi(\tau)$. It generalizes the BJP to include finite $x$ corrections and incorporates the full statistics of the random number of jumps through the moments $\langle n_t^k\rangle$, related to the distribution $Q_t(n)$. We also calculate the explicit perturbative expansion of the propagator $P(x,t)$ of a CTRW with exponentially-distributed waiting times.

\section{Perturbative expansion of the Big Jump Principle}
\label{Section3}

The geometric formulation of the convolution integral introduced in Section~\ref{Section2} allows us to express the conditional propagator $\phi(x|n)$ in terms of \eqref{g_y_n}, whose the $n$ cusps encode the $n$ possible realizations of the big jump event. In the BJP regime, each cusp corresponds to a configuration in which one jump is of order $x$, while the remaining $n-1$ jumps remain typically small. For stretched-exponential tails (i.e. $\beta<1$), configurations with more than one jump of order $x$ are exponentially suppressed, so the single big jump scenario provides the dominant contribution, with multiple large jumps contributing only subleading corrections. The zeroth-order contribution of each cusp yields the BJP scaling $\phi(x|n)\sim n f(x)$. For convenience, we build the perturbative series around the central cusp at $\vec y=0$. By symmetry of \eqref{g_y_n}, all cusps are equivalent, so the same construction can be reproduced around any of them (see Appendix~\ref{AppendixA}).

In this section we derive perturbative corrections to the leading BJP behavior of $\phi(x|n)$ in the large $x$ limit. We first treat the case $n=2$, where the structure is simplest, and then extend the construction to arbitrary $n$. The corrections originate from the collective contribution of the $n-1$ non-dominant summands. In the general case for arbitrary $n$, mixed terms arise in the multinomial expansion of $g(\vec y)$, corresponding to correlated fluctuations of several non-dominant jumps. Although present, their scaling with $x$ is always subleading compared to single jump contributions, and they do not affect the leading structure of the perturbative series.

Finally, we discuss the convergence limits and the asymptotic nature of the expansion. As is typical for asymptotic series, convergence at arbitrary order is not expected. To obtain controlled predictions at finite $x$, we introduce an analytical truncation procedure, which we call the optimal truncation method, which selects the order minimizing the truncation error.

\subsection{Perturbative expansion for $n=2$}

For two jumps the convolution can be written as
\begin{equation}
    \phi(x|2)
    = \tilde{N}^{2} |x|
      \int_{-\infty}^{\infty} dy\,
      \exp\!\left[-\alpha^\beta |x|^\beta g(y)\right],
    \qquad
    g(y)=|y|^\beta+|1-y|^\beta.
\end{equation}
In the BJP regime \(x\to\infty\), the integral is dominated by a neighbourhood of the cusp at \(y=0\), corresponding to the first jump being the big one and the second one remaining small. Expanding \(g(y)\) for \(|y|<1\), we obtain
\begin{equation}
    g(y)
    = |y|^\beta + 1
      + \sum_{l=1}^{\infty}(-1)^l\binom{\beta}{l}y^l,
    \label{eq:g_expansion_n2}
\end{equation}
where the condition \(|y|<1\) reflects the fact that the perturbative expansion is valid only when the small jump $\chi=xy$ is indeed small compared to the total displacement \(x\).  Substituting ~\eqref{eq:g_expansion_n2} into the convolution yields:
\begin{equation}
\label{phi_x_2_exponential_expansion}
 \phi(x|2) \approx 2\widetilde{N}^2 |x| \int_{-\infty}^{\infty} {\rm e}^{\alpha^\beta|x|^\beta (-1-|y|^\beta)}{\rm exp}\left[-\alpha^\beta|x|^\beta\sum_{l=1}^{\infty}(-1)^l\binom{\beta}{l}y^l\right]dy,
\end{equation}

Now, expanding the exponential and using the symmetry of the integrand, only even powers of \(y\) contribute, and the result can be written in the compact form
\begin{equation}
\label{phi_x_2_series}
    \phi(x|2)
    \sim 2 f(x)
    \left[
        1 + \sum_{\text{$l>0$, $l$ even}}^{\infty}
        \frac{d_l(x)}{x^l}\mathbb{E}[x^l]
    \right] \equiv 2f(x)G(x),
    \qquad x\to\infty,
\end{equation}

where $d_l(x)$ are the coefficients arising from the expansion of the exponential, \(\mathbb{E}[x^l]\) denotes the \(l\)-th moment of the PDF \eqref{f_x}. The first few coefficients are reported in Table \ref{tab:dl_coeffs_full}.
\begin{table}[p]
\centering

\setlength{\tabcolsep}{3pt}
\renewcommand{\arraystretch}{1.15}
\tiny

\begin{tabular}{c l}
\hline\hline
$l$ & \text{perturbative coefficient $d_l(x)$}\\
\hline

$2$ &
\parbox[t]{0.92\textwidth}{$
d_2(x)=\frac{\beta}{2}\,\alpha^{\beta}|x|^{\beta}
\Big(\beta\,\alpha^{\beta}|x|^{\beta}-\beta+1\Big)
$}
\\[6pt]

$4$ &
\parbox[t]{0.92\textwidth}{$\displaystyle
\begin{aligned}
d_4(x)=\frac{\beta}{24}\,\alpha^{\beta}|x|^{\beta}\Big(
&\beta^3\alpha^{3\beta}|x|^{3\beta}
-6\beta^3\alpha^{2\beta}|x|^{2\beta}
+7\beta^3\alpha^{\beta}|x|^{\beta}
-\beta^3 \\
&+6\beta^2\alpha^{2\beta}|x|^{2\beta}
-18\beta^2\alpha^{\beta}|x|^{\beta}
+6\beta^2
+11\beta\alpha^{\beta}|x|^{\beta}
-11\beta
+6
\Big)
\end{aligned}
$}
\\[8pt]

$6$ &
\parbox[t]{0.92\textwidth}{$\displaystyle
\begin{aligned}
d_6(x)=\frac{\beta}{720}\,\alpha^{\beta}|x|^{\beta}\Big(
&\beta^5\alpha^{5\beta}|x|^{5\beta}
-15\beta^5\alpha^{4\beta}|x|^{4\beta}
+65\beta^5\alpha^{3\beta}|x|^{3\beta}
-90\beta^5\alpha^{2\beta}|x|^{2\beta}
+31\beta^5\alpha^{\beta}|x|^{\beta}
-\beta^5 \\
&+15\beta^4\alpha^{4\beta}|x|^{4\beta}
-150\beta^4\alpha^{3\beta}|x|^{3\beta}
+375\beta^4\alpha^{2\beta}|x|^{2\beta}
-225\beta^4\alpha^{\beta}|x|^{\beta}
+15\beta^4 \\
&+85\beta^3\alpha^{3\beta}|x|^{3\beta}
-510\beta^3\alpha^{2\beta}|x|^{2\beta}
+595\beta^3\alpha^{\beta}|x|^{\beta}
-85\beta^3 \\
&+225\beta^2\alpha^{2\beta}|x|^{2\beta}
-675\beta^2\alpha^{\beta}|x|^{\beta}
+225\beta^2 \\
&+274\beta\alpha^{\beta}|x|^{\beta}
-274\beta
+120
\Big)
\end{aligned}
$}
\\[8pt]

$8$ &
\parbox[t]{0.92\textwidth}{$\displaystyle
\begin{aligned}
d_8(x)=\frac{\beta}{40320}\,\alpha^{\beta}|x|^{\beta}\Big(
&\beta^7\alpha^{7\beta}|x|^{7\beta}
-28\beta^7\alpha^{6\beta}|x|^{6\beta}
+266\beta^7\alpha^{5\beta}|x|^{5\beta}
-1050\beta^7\alpha^{4\beta}|x|^{4\beta} \\
&\quad +1701\beta^7\alpha^{3\beta}|x|^{3\beta}
-966\beta^7\alpha^{2\beta}|x|^{2\beta}
+127\beta^7\alpha^{\beta}|x|^{\beta}
-\beta^7 \\
&+28\beta^6\alpha^{6\beta}|x|^{6\beta}
-588\beta^6\alpha^{5\beta}|x|^{5\beta}
+3920\beta^6\alpha^{4\beta}|x|^{4\beta}
-9800\beta^6\alpha^{3\beta}|x|^{3\beta} \\
&\quad +8428\beta^6\alpha^{2\beta}|x|^{2\beta}
-1764\beta^6\alpha^{\beta}|x|^{\beta}
+28\beta^6 \\
&+322\beta^5\alpha^{5\beta}|x|^{5\beta}
-4830\beta^5\alpha^{4\beta}|x|^{4\beta}
+20930\beta^5\alpha^{3\beta}|x|^{3\beta}
-28980\beta^5\alpha^{2\beta}|x|^{2\beta} \\
&\quad +9982\beta^5\alpha^{\beta}|x|^{\beta}
-322\beta^5 \\
&+1960\beta^4\alpha^{4\beta}|x|^{4\beta}
-19600\beta^4\alpha^{3\beta}|x|^{3\beta}
+49000\beta^4\alpha^{2\beta}|x|^{2\beta}
-29400\beta^4\alpha^{\beta}|x|^{\beta}
+1960\beta^4 \\
&+6769\beta^3\alpha^{3\beta}|x|^{3\beta}
-40614\beta^3\alpha^{2\beta}|x|^{2\beta}
+47383\beta^3\alpha^{\beta}|x|^{\beta}
-6769\beta^3 \\
&+13132\beta^2\alpha^{2\beta}|x|^{2\beta}
-39396\beta^2\alpha^{\beta}|x|^{\beta}
+13132\beta^2 \\
&+13068\beta\alpha^{\beta}|x|^{\beta}
-13068\beta
+5040
\Big)
\end{aligned}
$}
\\

\hline\hline
\end{tabular}

\normalsize
\caption{Explicit expressions of the perturbative coefficients $d_l(x)$ obtained from Eq.(\ref{d_l_x_general}) for the perturbative orders $l=2,4,6,8$. Notice that $d_2(x)$ is exactly the same first order correction found in \cite{Omer_ref}. Note that the perturbative ordering of the series expansion \eqref{phi_x_n_pert} is controlled by the combinations $d_l(x)/x^l$ entering $G(x)$, rather than by $d_l(x)$ alone. For $0<\beta<1$, the first perturbative term $d_2(x)$ for example contains both $|x|^{2\beta}$ and $|x|^{\beta}$ contributions, but in $d_2(x)/x^2$ these scale as $|x|^{2\beta-2}$ and $|x|^{\beta-2}$, respectively, so the latter is subleading as $|x|\to\infty$. Similar subleading powers appear at higher orders and can be kept or neglected depending on the desired accuracy.}
\label{tab:dl_coeffs_full}
\end{table}
In this work, instead of plotting $G(x)$, which requires the calculation of an infinite number of terms, we will use a truncated sum, denoted as 
\begin{eqnarray}
    G^{(L)}(x)\equiv 1 + \sum_{\text{$l>0$, $l$ even}}^{L}
        \frac{d_l(x)}{x^l}\mathbb{E}[x^l],
\end{eqnarray}
with $G^{(0)}(x)=1$, corresponding to the single big jump. Hence, the $L$-th perturbative expansion to the BJP limit of $\phi(x|2)$ reads $\phi(x|2) \approx 2 f(x)\left(1+ G^{(L)}(x)\right)$,
which accurately captures its deviations from the asymptotic regime even when \(x\) is moderately large, and in particular when \(\beta\to1^{-}\), where convergence to the leading BJP term is slow.

\subsection{Generalization to arbitrary $n$}
\label{Subsection3B}

For a generic number of jumps \(n\), the convolution takes the form reported in Eq. \eqref{convolution},
with \(g(\vec y)\) defined in Eq. \eqref{g_y_n}. Expanding \(g(\vec y)\) around the central cusp at \(\vec y=\vec 0\), we obtain
\begin{equation}
\label{g_y_expansion_n}
    g(\vec y)
    = \sum_{i=1}^{n-1}|y_i|^\beta + 1
      + \sum_{l=1}^{\infty}(-1)^l\binom{\beta}{l}
        \left(\sum_{i=1}^{n-1} y_i\right)^l.
\end{equation}

The last term of \eqref{g_y_expansion_n} can be expanded using the Newton multinomial theorem \cite{Multinomial_thm} as:
\begin{equation}
\label{multinomial_dev}
    \left(\sum_{i=1}^{n-1} y_i\right)^l
    = \sum_{l_1+\cdots+l_m=l}
      \frac{l!}{l_1!\cdots l_m!}
      y_{i_1}^{l_1}\cdots y_{i_m}^{l_m},
\end{equation}
revealing two types of contributions:

\begin{itemize}
    \item \emph{Coherent terms}, of the form \(y_i^l\), describing independent fluctuations of each of the \(n-1\) small jumps;
    \item \emph{Mixed terms}, of the form \(y_{i_1}^{l_1}y_{i_2}^{l_2}\cdots y_{i_m}^{l_m}\) with \(m\ge2\), which couple fluctuations of different small jumps;
\end{itemize}

Physically, the perturbative construction is organized around configurations in which a single displacement is of order \(x\), while the remaining \(n-1\) jumps are collectively subdominant. For stretched-exponential tails, i.e. \(\beta<1\), configurations containing more than one jump of order \(x\) are exponentially suppressed and therefore do not contribute to the leading saddle-point structure. Mixed terms instead describe correlated fluctuations of several non-dominant jumps. These contributions are not discarded a priori; rather, their scaling with \(x\) is systematically smaller than that of the coherent terms, and their cumulative effect produces only subleading corrections to the large-\(x\) asymptotics. 
Considering only the coherent terms, the integral in Eq. \eqref{convolution} factorizes into \(n-1\) identical one-dimensional integrals, and we obtain:
\begin{equation}
    \phi(x|n)
    \sim n f(x)\,[G(x)]^{n-1},
    \qquad |x|\to\infty,
    \label{phi_general_n_G0}
\end{equation}
where the function $G(x)$ is the previous one defined in \eqref{phi_x_n_pert} and contains all the information about the perturbative corrections. Importantly, $G(x)$ depends only on the large $x$ structure of the expansion and we remember that it is independent of $n$. 
The entire derivation of $\phi(x|n)$ in which mixed terms are included, together with scaling arguments and numerical checks establishing their sub-leading role, is presented in Appendix \ref{AppendixC}. Equation ~\eqref{phi_general_n_G0} holds for arbitrary fixed $n$ and does not rely on a large $n$ limit, in contrast to standard large deviation approaches. This is exactly the result introduced in Section \ref{Section2}, and it represents the full perturbative generalization of the distribution of the sum of $n$ IID RVs with stretched-exponential PDF. 

\begin{figure}[ht]
\centering
\includegraphics[width=\textwidth]{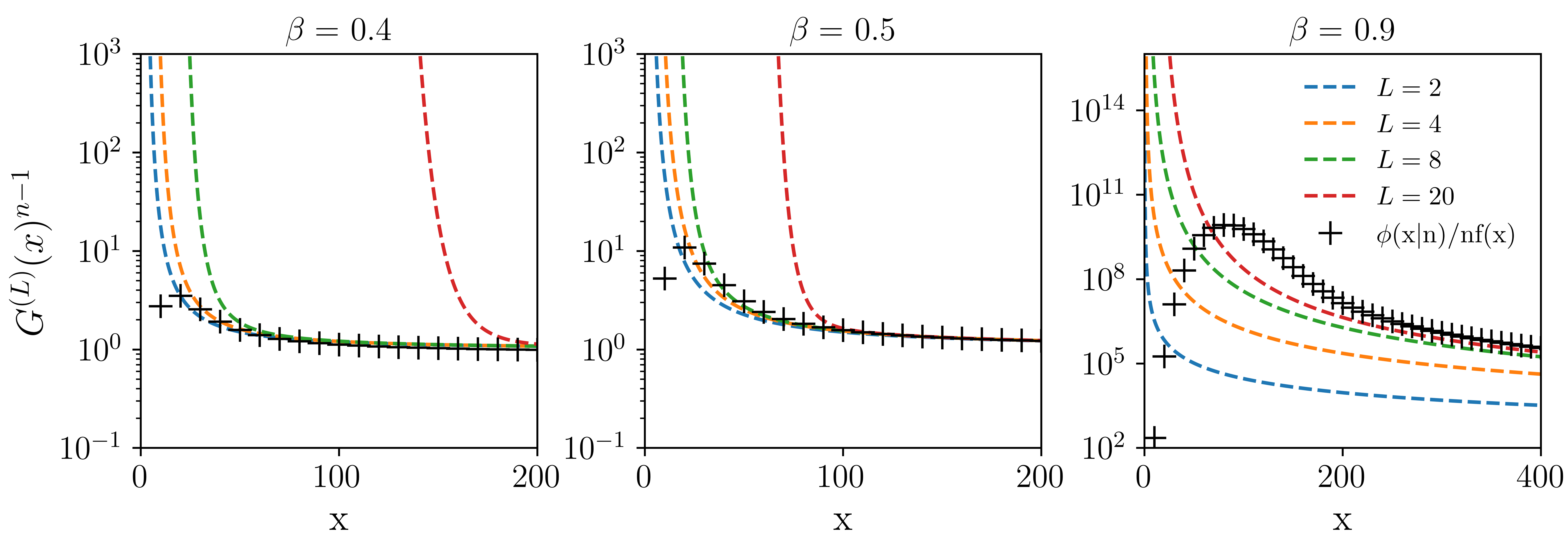}
\caption{Plots of our perturbative correction obtained in Eq.\eqref{phi_general_n_G0} (dashed colored lines) compared to the numerical results (black crosses), for $\beta = 0.4$ (left), $\beta = 0.5$ (middle), and $\beta = 0.9$ (right). The number of jumps is set to $n=30$, and the plots are in log-scale and normalized with respect to the BJP asymptotic limit $nf(x)$. While for any $\beta<1$ and $x\to\infty$ the BJP leading term converges to the numerical results, for finite values of $x$ and incresasing $\beta$ the higher order correction terms becomes more relevant. The labels $L$ means a $L$-th truncated perturbative series, i.e. $G^{(L)}(x)$. In these plots we use $L=2, 4, 8, 20$.} \label{Figure2}
\end{figure}

In Figure \ref{Figure2} we compare truncated perturbative expansions with numerical simulations. The plotted quantity is normalized with respect to the BJP asymptotic limit $n f(x)$, so that deviations from the leading big jump behavior are highlighted. Increasing the perturbative order $L$ improves the agreement, particularly as $\beta$ approaches $1^{-}$, where convergence to the BJP becomes slower and higher-order corrections are more relevant.

\subsection{Divergence and optimal truncation method}
\label{Subsection3C}

The series defining \(G(x)\) in Eq. \eqref{phi_x_n_pert} is asymptotic rather than convergent. This behaviour originates from the domain of validity of the Taylor expansion of $g(\vec{y})$: Eq. \eqref{g_y_expansion_n} holds only when \(|y_i|<1\) for every $i$, whereas the convolution integrals in Eq. \eqref{convolution} extend over the full domain \(\mathbb{R}^{n-1}\). For large but finite \(x\), the exponential factor strongly suppresses regions where \(|y_i|>1\), which justifies the use of the expansion at low perturbative orders. However, as the perturbative order increases at fixed \(x\), the series progressively probes configurations where the local expansion of \(g(\vec y)\) is no longer accurate, and the series eventually diverges. This is a standard feature of asymptotic expansions and does not signal a pathology of the method. Many widely used perturbative approximations in probability theory and statistical physics share this structure; in particular, the classical Edgeworth expansion is also asymptotic rather than convergent, yet provides highly accurate approximations when truncated optimally. As in that context, the present series is used as a controlled practical approximation rather than a convergent representation \cite{Asymptotic_series}.

The onset of divergence is directly visible in the first plot of Figure \ref{Figure2}, where the expansion truncated at high perturbative order begins to deviate from numerical results at moderate values of $x$. This behaviour calls for an optimal truncation prescription. We write the truncated expansion of $G(x)$ as
\begin{equation}
\label{G_0_g_l}
    G^{(L)}(x)
    = 1 + \sum_{\text{$L$ even}}^{L} \mathcal{G}^{(L)}(x),
    \qquad
    \mathcal{G}^{(L)}(x) = \frac{d_L(x)}{x^L}\,\mathbb{E}[x^L].
\end{equation}

Numerically, for fixed \(x\), the sequence \(\{\mathcal{G}^{(L)}(x)\}\) initially decreases with increasing perturbative order, reaches a minimum at some order \(L^*(x)\), and then grows. The optimal truncation order at fixed $x$ is therefore defined as
\begin{equation}
    L^*(x)
    = \arg\min_{L} \left|\mathcal{G}^{(L)}(x)\right|.
\end{equation}

The optimally truncated approximation of $\phi(x|n)$ is obtained by summing the series up to $L^*(x)$, where $L^*(x)$ is a positive integer. Physically, low-order terms describe the dominant fluctuations of the small jumps around the BJP regime, while very high-order terms encode rare excursions of the variables \(y_i\) toward \(|y_i|\sim 1\) or larger, where the local expansion of \(g(\vec y)\) ceases to be accurate. Truncation at the minimal term ensures that the expansion is evaluated within the region where it faithfully represents the local geometry of the cusp.

The behaviour of the optimal truncation order \(L^*(x)\) depends on \(\beta\). For relatively small \(\beta\), the cusps of \(g(\vec y)\) are sharp and the big jump mechanism dominates already at modest values of \(x\); in this case only a few correction terms are required. In contrast, as \(\beta \to 1^{-}\), the cusp becomes shallower and convergence to the BJP slows down; the optimal truncation order increases and more terms are needed to resolve the intermediate regime accurately. As illustrated numerically in Figure \ref{Figure4}, the optimally truncated series, denoted \(G^{(L^*)}(x)\), provides an accurate approximation to \(\phi(x|n)\), matching the exact numerical convolution over several decades in \(x\), reproducing the correct asymptotic behaviour for \(x\to\infty\), and extending the predictive power of the theory deep into the crossover region between the CLT and BJP regimes. The optimally truncated contribution \(G^{(L^*)}(x)\) is the key ingredient entering the perturbative expression for the CTRW propagator \(P(x,t)\) discussed in Section \ref{Section5}.

\begin{figure}[ht]
\centering
\includegraphics[width=\textwidth]{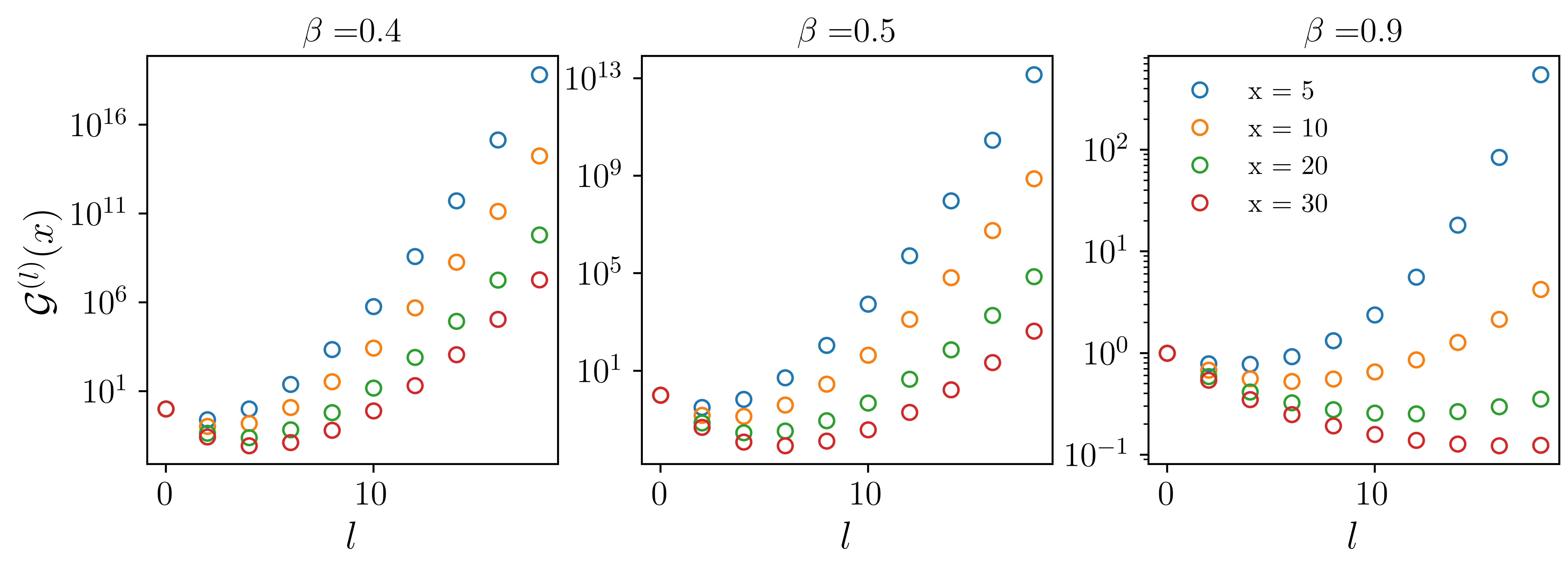}
\caption{Location of the minimal term $\mathcal{G}^{(L^*)}(x)$ of the perturbative series $G(x)$ as a function of $x$ and $\beta$. As $x$ increases, a larger number of perturbative terms remain non-divergent, and the optimal truncation order $L^*(x)$ grows. For moderate $x$, only a few correction terms should be retained. As $\beta\to1^{-}$, convergence to the BJP becomes slower and more terms are required. The plots are shown in log-scale for $x=5, 10, 20, 30$, with $\beta=0.4$ (left), $\beta=0.5$ (middle) and $\beta=0.9$ (right).}\label{Figure3}
\end{figure}
\begin{figure}[ht]
\centering
\includegraphics[width=\textwidth]{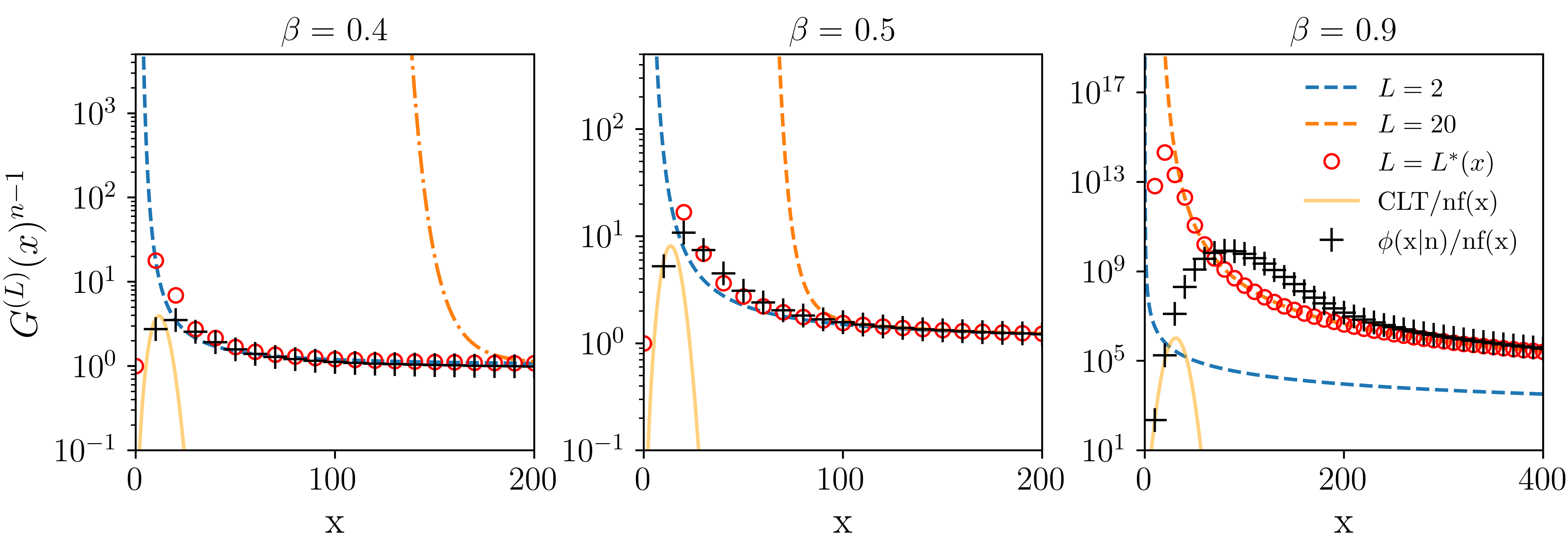}
\caption{
Conditional PDF $\phi(x|n)$ normalized by the BJP asymptotic behavior $n f(x)$ (black crosses). Dashed colored curves show perturbative approximations from Eq.~\eqref{phi_general_n_G0} truncated at orders $L=2$ and $L=20$. The orange line shows the CLT prediction from Eq.~\eqref{phi_x_n_clt}. Red circles represent the optimally truncated theory using $G^{(L^*)}(x)$. Simulations are shown in log-scale for $n=30$, with $\beta=0.4$ (left), $\beta=0.5$ (middle) and $\beta=0.9$ (right). The optimal truncation provides excellent agreement with numerical results even for moderate $x$.}\label{Figure4}
\end{figure}

\section{Rate function}
\label{Section4}

In this section we show that the leading contributions of the perturbative expansion for $\phi(x|n)$ in Eq.~\eqref{phi_x_n_pert}, derived in Sec.~\ref{Section3}, encode the anomalous scaling that characterizes the crossover between the Gaussian regime described by the CLT and the BJP tail regime. In particular, in the first subsection we show that the leading contribution of the first perturbative correction to the BJP, $d_2(x)$, naturally yields the scaling behaviour predicted by large deviation theory for stretched-exponential distributions \cite{overdamped_OU_4}. While the corresponding scaling exponents can be inferred from matching arguments between the CLT and the BJP asymptotics, here they emerge directly from the structure of the large $x$ perturbative expansion, without invoking the Gaussian regime. In the second subsection we show that this contribution leads to an analytical expression for an approximate rate function $\mathcal{I}_{approx}(x)$, consistent with recent results obtained from the complementary large $n$ perturbative approach \cite{LDT_Perturbative_Naftali}.

\subsection{Anomalous scaling and critical exponents}

We restart from the perturbative expansion of $\phi(x|n)$ \eqref{phi_x_n_pert} derived in previous section. Starting with a simple scaling analysis of the general expression of $d_l(x)$, i.e. Eq. \eqref{d_l_x_general_main}, in the limit of $x \rightarrow \infty$, the $L$-th order leading term for $\beta<1$ goes as:
\begin{equation}
    \frac{d_L(x)}{x^L} \underset{x \rightarrow \infty}{\sim}  c_L |x|^{L(\beta-1)}
\end{equation}
Here $c_L$ is a constant pre-factor independent of $x$, and so the leading term of every $L$-th perturbative contribution scales as $|x|^{L(\beta-1)}$. Higher perturbative orders therefore decay faster with $x$, and the dominant contribution to the perturbative series in the large $x$ limit is controlled by the lowest nontrivial order $L=2$. Higher orders (e.g. $L=4,6,\dots$) provide subleading corrections that modify prefactors and higher-order scaling terms but do not affect the leading power-law structure. For $\beta<1$ the leading behavior of the perturbative series is therefore encoded in $L=2$, i.e. $d_2(x)$, reported in Table \ref{tab:dl_coeffs_full}. The leading behavior of the perturbative series is then:

\begin{equation}
    \frac{d_2(x)}{x^2} \underset{x \rightarrow \infty}{\sim} \frac{1}{2} \beta^2 \alpha^{2\beta} |x|^{2\beta-2}
\end{equation}

Thus, the leading first order correction for the PDF $\phi(x|n)$ in the large $x$ regime and for a finite number of jumps $n$ reads as:

\begin{equation}
\phi(x|n) \underset{x \rightarrow \infty}{\sim} n f(x) \left( 1 + \frac{1}{2} \beta^2 \alpha^{2\beta} |x|^{2\beta - 2} \right)^{n-1}
\end{equation}

Exponentiating the equation above in large $x$ limit and neglecting the $n$-independent terms since they give only subleading logarithmic corrections for large $x$, $\phi(x|n)$ scales as:

\begin{align}
\label{scaling_phi_x}
\phi(x|n)
& \underset{x\rightarrow \infty}{\sim} \exp \bigg\{ -\alpha^{\beta}|x|^{\beta} + \frac{n}{2} \beta^2 \alpha^{2\beta} |x|^{2\beta - 2} \bigg\}
\end{align} 

At this stage the expression is derived for large $x$ at fixed $n$. To extract the crossover scaling with $n$, we now introduce a generalized anomalous large deviation ansatz following \cite{overdamped_OU_4, LDT_Perturbative_Naftali}:

\begin{equation}
-\frac{\log \phi(x|n)}{n^{\eta}} \underset{\substack{x \rightarrow \infty \\ \text{$x/n^{\xi}$ fixed}}}{\asymp}  \mathcal{I} \left( \frac{|x|}{n^{\xi}} \right) 
\end{equation}

Where $\eta$ and $\xi$ are two anomalous scaling exponents that describe the typical scale of $|x|$ in terms of $n$, i.e. the typical jump length $\ell(n)$ scales with the number of jumps $n$ as $\ell(n)\sim n^{\xi}$. The scaling function $\mathcal{I}(.)$ is a generalized rate function. Inserting our large deviation ansatz into \eqref{scaling_phi_x}, we obtain consistency conditions that fix the scaling exponents $\xi$ and $\eta$. We finally obtain:

\begin{equation}
    \xi=\frac{1}{2-\beta} , \qquad \eta=\beta \xi = \frac{\beta}{2-\beta} 
\end{equation}

Both anomalous exponents $\eta, \xi < 1$ change continuously with $\beta$, and as $\beta \rightarrow 1$ we recover the usual large deviation principle, since $\eta=\xi=1$, corresponding to exponential tails. Since $\beta<1$ the tail behavior is slower than exponential, and for the typical scale length $\ell(n) \sim n^{\frac{1}{2-\beta}}$ the CLT contributions in the bulk are of the same order as the BJP events in the tails of the distribution $\phi(x|n)$. This signals the presence of a dynamical phase transition, as rigorously demonstrated in \cite{LDT_3, LDT_4, LDT_5} in the large $n$ limit. Near the origin, however, the generalized rate function reduces to the standard quadratic form, reflecting Gaussian fluctuations in agreement with the CLT. Thus the full rate function interpolates between a parabolic structure at small deviations and the anomalous scaling described above in the far-tail regime. As a final remark, the anomalous exponents $\eta$ and $\xi$ coincide with those obtained in \cite{overdamped_OU_4, LDT_Perturbative_Naftali}, but here they emerge directly from the large $x$ perturbative structure.

\subsection{Approximated rate function}

Thus, we found that in the limit of large $x$ and large $n$ with the ratio $x/n^{\xi}$ fixed, the contribution of the BJP regime and its leading corrective term coming from the perturbative expansion of $\phi(x|n)$ \eqref{phi_x_n_pert} follows an anomalous large deviation scaling of the form:
\begin{equation}
-\frac{\log \phi(x|n)}{n^{\frac{\beta}{2-\beta}}} \underset{\substack{x \rightarrow \infty \\ \text{$x/n^{\xi}$ fixed}}}{\asymp}  \mathcal{I}_{approx} \left( r=\frac{|x|}{n^{\frac{1}{2-\beta}}} \right) .
\end{equation}
Here $\mathcal{I}_{approx}(r)$ is an approximate rate function obtained from the large $x$ perturbative structure at fixed $n$. It approximates the full rate function derived within large deviation theory in the limit $n\to\infty$ \cite{overdamped_OU_4}. It reads:
\begin{equation}
\label{I_approx}
\mathcal{I}_{approx}(r)
= \alpha^{\beta}|r|^{\beta}
- \frac{1}{2}\alpha^{2\beta}\beta^2 |r|^{2\beta-2}.
\end{equation}
The first term corresponds to the cost of producing a single big jump of length $x$ after $n$ jumps, as prescribed by the BJP, while the second term originates from the collective Gaussian fluctuations of the remaining $n-1$ small jumps. Its dependence $\sim |r|^{2\beta-2}$ becomes increasingly relevant when convergence to the BJP becomes slow, i.e. when $\beta \rightarrow 1^{-}$.

To test the consistency of Eq.~\eqref{I_approx}, we compare it with the full rate function obtained from large–deviation theory for IID stretched–exponential random variables \cite{LDT_3, LDT_4, LDT_5}:
\begin{equation}
\label{I_exact}
\mathcal{I}(r)
=
\min_{t\in[0,r]}
\left\{
\alpha^{\beta}\,|t|^{\beta}
+
\frac{1}{2}\,|r-t|^{2}
\right\}.
\end{equation}
We now analyze the large $r$ regime of this expression. For $r\to\infty$, the minimizer satisfies $t\simeq r$, and we write $t=r(1-\varepsilon)$ with $\varepsilon\ll1$. Minimizing Eq.~\eqref{I_exact} and solving for $\varepsilon$ gives, to leading order:
\begin{equation}
\varepsilon
\simeq 
\beta\,\alpha^{\beta}\,|r|^{\beta-2}.
\end{equation}
Substituting this expression into Eq.~\eqref{I_exact} and expanding for large $r$ yields:
\begin{equation}
\label{eq:I_asymptotic}
\mathcal{I}(r)
\simeq
\alpha^{\beta}|r|^{\beta}
-\frac{1}{2}\,\beta^{2}\,\alpha^{2\beta}\,|r|^{\,2\beta-2}
+ O(|r|^{\,3\beta-4}),
\qquad r\to\infty.
\end{equation}

\begin{figure}[ht]
\centering
\includegraphics[width=0.90\textwidth]{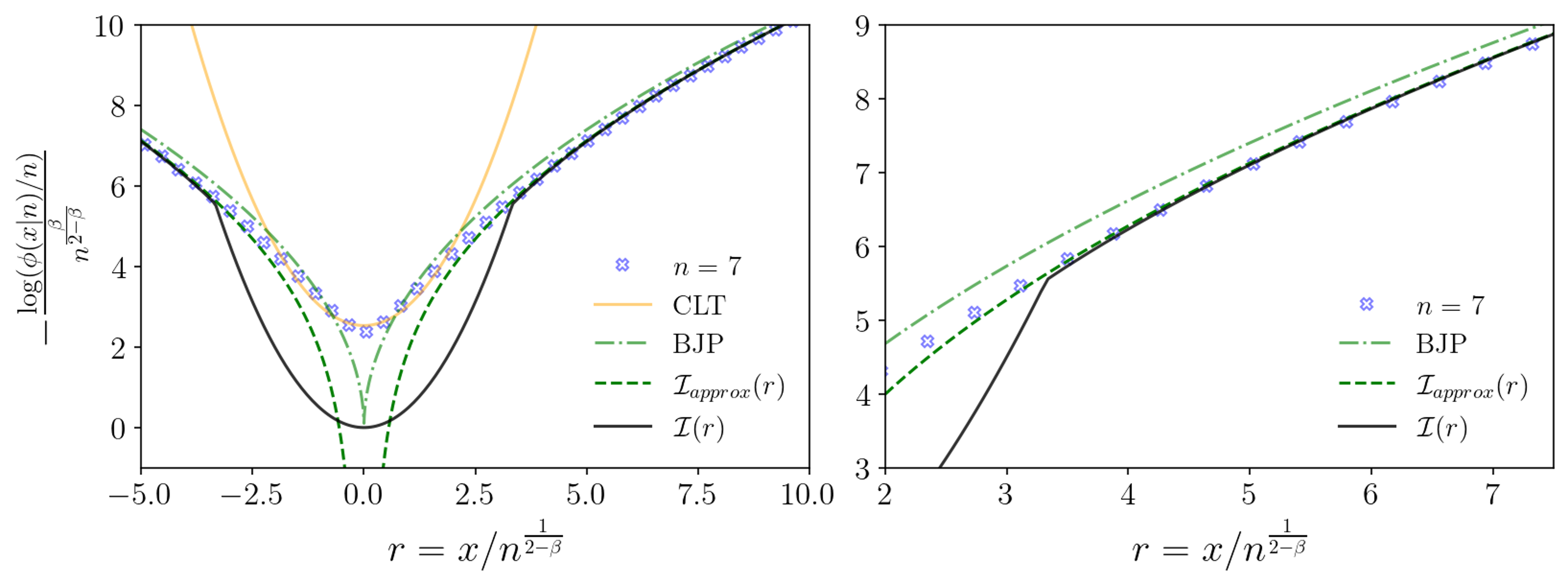}
\includegraphics[width=0.90\textwidth]{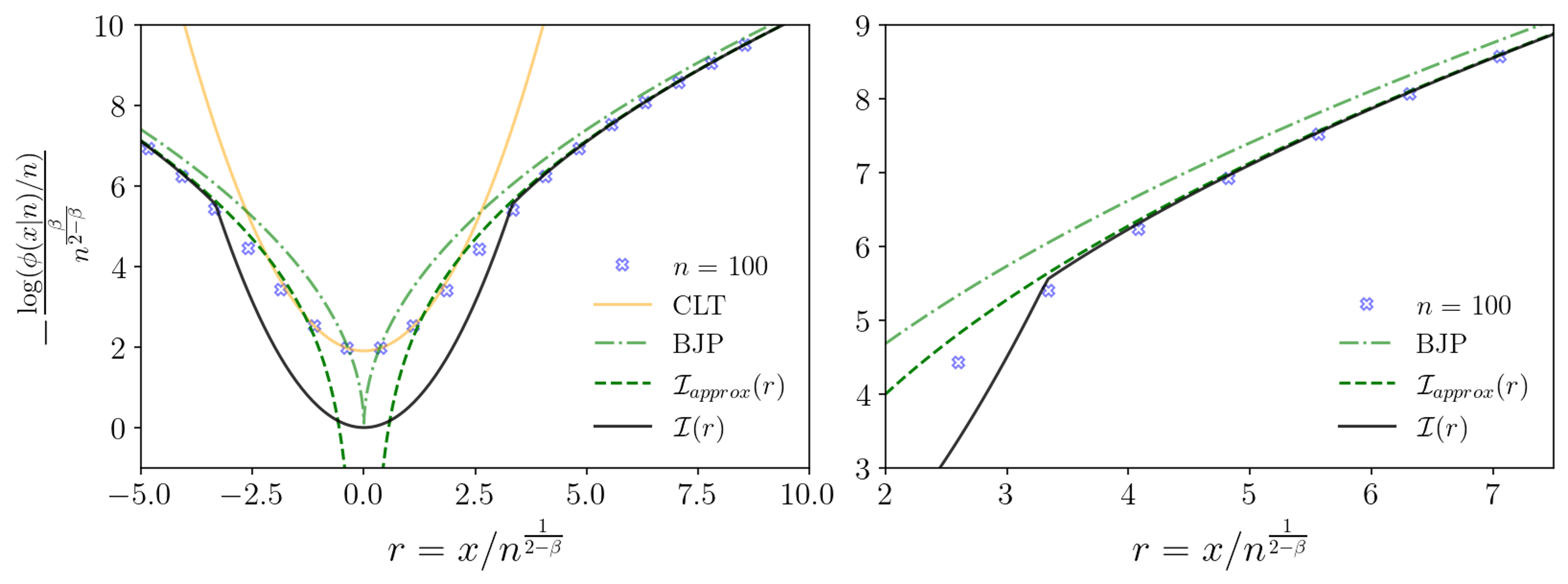}
\caption{Numerical simulations of the sum of IID stretched-exponential random variables distributed according to Eq.~\eqref{f_x}. Results are shown for $\beta=0.5$ and two values of the number of summands, $n=7$ and $n=100$. Blue crosses represent the empirical rate function extracted from simulations. The yellow parabola shows the quadratic rate function associated with typical Gaussian fluctuations predicted by the CLT. The black curve shows the full rate function given by Eq.~\eqref{I_exact}, obtained from large-deviation theory in the limit $n\to\infty$. The green dotted curve shows the leading BJP contribution $\alpha^{\beta}|r|^{\beta}$, while the green dashed curve shows the large $r$ approximation $\mathcal{I}_{approx}(r)$ from Eq.~\eqref{I_approx} derived from the perturbative expansion. For moderate $n$, the numerical data interpolate smoothly between the Gaussian regime and the large $r$ behaviour. The perturbative large $r$ approximation captures the behaviour of the distribution in the big jump regime and provides a good description of moderate deviations, even away from the strict large deviation limit.}
\label{Figure5}
\end{figure}

This coincides with Eq.~\eqref{I_approx}. Therefore, the leading non BJP contribution obtained from the perturbative coefficient $d_2(x)$ reproduces the leading large $r$ expansion of the full rate function. Higher perturbative orders do not modify this leading scaling structure. They generate subleading corrections in the large $x$ limit and correspond to finite $n$ refinements of the approximation. The perturbative expansion therefore provides a controlled large $x$ approximation of the rate function that is consistent with the asymptotic structure predicted by large-deviation theory. 

In Fig.~\ref{Figure5} we compare numerical simulations of sums of IID stretched-exponential variables with the full rate function and with its large $r$ approximation obtained from the perturbative expansion. The non-analyticity of the full rate function, which signals the dynamical phase transition in the strict large deviation limit, emerges only when $n\to\infty$. Since the perturbative expansion developed here is performed at finite $n$ and in the large $x$ regime, it captures the large $r$ asymptotic structure of the rate function but does not show the non-analytic feature associated with the phase transition itself.

\newpage

\section{CTRW}
\label{Section5}

In this section we return to the CTRW model introduced in Section \ref{Section2} and combine the perturbative description of $\phi(x|n)$ derived in Section \ref{Section3} with the subordination relation in order to obtain an approximation for the propagator $P(x,t)$ in the large $x$ limit. The key difference with respect to the previous sections is that the number of jumps $n$ is now a random variable: for a given realization, the walker performs a random number of jumps $n$ up to time $t$, and remember that we denote by $Q_t(n)$ as the probability of having exactly $n$ jumps at time $t$. Infact, the propagator $P(x,t)$ is governed by the usual subordination relation \eqref{subordination} that assumes no correlation between the temporal RVs associated with the waiting times PDF $\psi(\tau)$ and the spacial RVs associated with the jump length distribution $f(\chi)$ \eqref{f_x}. We will start in the first subsection deriving the general perturbative expression of $P(x,t)$ \eqref{P_x_t_pert}, then in the second subsection we will explore the particular case of an exponential distribution of the waiting times PDF $\psi(\tau)$. 

\subsection{General perturbative expansion of the CTRW}

From Sec.~\ref{Section3} we have obtained an analytical series expansion for $\phi(x|n)$ \eqref{phi_x_n_pert} that, neglecting mixed terms and focusing on the large $x$ regime, encodes the full perturbative structure in powers of $1/x^2$. Inserting directly Eq. \eqref{phi_x_n_pert} in the subordination relation for CTRWs \eqref{subordination} yields
\begin{equation}
\label{P_x_t_start}
P(x,t)
= f(x)\sum_{n=0}^{\infty} Q_t(n)\,n\,[G(x)]^{\,n-1}.
\end{equation}
Since we are interested in large $x$ limit, is convenient to define the small parameter $\Delta G(x) = G(x)-1$
and expand the $n-1$-th power, since for $x\rightarrow \infty$ we have that $\Delta G(x) \rightarrow 0$. It yields:
\begin{equation}
\big[G(x)\big]^{\,n-1}
=\big[1+\Delta G(x)\big]^{n-1}
=1+\sum_{m=1}^{\infty}\binom{n-1}{m}\,[\Delta G(x)]^{m}.
\end{equation}
Now, focusing on the factorial, we can use a common identity to express it as a sum of polynomials in $n$:
\begin{equation}
\binom{n-1}{m}
=\frac{1}{m!}\sum_{k=0}^{m}(-1)^k\binom{m}{k}\,n^{\,m-k}.
\end{equation}
So, rewriting the binomial coefficient as a polynomial in $n$ and substituting it back into Eq.~\eqref{P_x_t_start} and exchanging the order of sums gives:
\begin{align}
P(x,t)
=
f(x)\left[
\sum_{n=0}^{\infty}nQ_t(n)
+ \sum_{m=1}^{\infty}\frac{[\Delta G(x)]^{m}}{m!}
\sum_{k=0}^{m}(-1)^k\binom{m}{k}
\sum_{n=0}^{\infty}n^{\,m-k+1} Q_t(n)
\right].
\end{align}
We now introduce the $p$-th ordered moments of the number $n$ of jumps in time $t$ distribution $Q_n(t)$, defined as $\langle n_t^p\rangle
=\sum_{n=0}^{\infty} n^{p} Q_t(n)$.
In terms of these moments, we obtain the final expression for the perturbative expansion of the propagator $P(x,t)$:
\begin{equation}
\label{P_x_t_general}
P(x,t)
\underset{|x|\to\infty}{\sim} f(x)\left[
\langle n_t\rangle
+ \sum_{m=1}^{\infty}\frac{[\Delta G(x)]^{m}}{m!}
\left(
\sum_{k=0}^{m}(-1)^k\binom{m}{k}\,\langle n_t^{\,m-k+1}\rangle
\right)
\right], \qquad \Delta  G(x)=G(x)-1.
\end{equation}

We finally derived formula \eqref{P_x_t_pert} shown in Section \ref{Section2}. The physical interpretation is now transparent. The leading term $\langle n_t\rangle f(x)$ is the BJP contribution: the probability of making a single jump of size $x$ times the mean number of attempts $\langle n_t\rangle$ in the total measurement time $t$ in which it can occur. The higher-order terms describe how the ensemble of small jumps around the big jump contribution modifies this picture; they are controlled by the higher $p$-th moments $\langle n_t^p\rangle$ of the number of jumps distribution $Q_n(t)$ and by the large $x$ corrections encoded in $\Delta G(x)$. Since $Q_n(t)$ and their moments are strictly related with the waiting time PDF $\psi(\tau)$, the only regularity condition that we need to have the $p$-th moments $\langle n_t^p\rangle$ finite at every order is to require finite average waiting time $\langle \tau \rangle$. 

\subsection{The exponential waiting time PDF case}

We now specialize Eq.~\eqref{P_x_t_general} to the case of exponential waiting times $\psi(\tau)=e^{-\tau}$, assuming for simplicity $\langle \tau \rangle=1$.
As a consequence, the distribution of the number $n$ of jumps $Q_n(t)$ during the total time $t$ follows a Poisson distribution with mean $\langle n_t\rangle =t$, that is:
\begin{equation}
Q_t(n)=\frac{t^n\,e^{-t}}{n!}.
\end{equation}
The $p$-th ordered moments $\langle n_t^p\rangle$ are well known in mathematical literature \cite{Poisson}, and could be expressed formally in terms of the Touchard polynomials $T_p(t)$,
\begin{equation}
\langle n_t^p\rangle
= T_p(t)
=\sum_{i=1}^{p}\stirling{p}{i}\,t^i,
\end{equation}
where $\stirling{p}{i} = \sum_{i=1}^p \left( \sum_{j=1}^i \frac{(-1)^{i-j}j^p}{(i-j)!j!} \right) t^i$ are Stirling numbers of the second kind. Inserting these moments into Eq.~\eqref{P_x_t_general}, we obtain the final perturbative expression for the propagator $P(x,t)$ of a CTRW with sub-exponential stretched jumps distribution and exponential waiting time distribution:
\begin{equation}
\label{P_x_t_poisson}
P(x,t)
\underset{|x|\to\infty}{\sim} f(x)\left[
t
+ \sum_{m=1}^{\infty}\frac{[G(x)-1]^{m}}{m!}
\left(
\sum_{k=0}^{m}(-1)^k\binom{m}{k}\,T_{m-k+1}(t)
\right)
\right].
\end{equation}
The first perturbative corrections at lower orders illustrate how the contribution of the small fluctuations behaves around the BJP result $P_{\mathrm{BJP}}(x,t)\sim t f(x)$ as a function of the total displacement $x$ and of the total measurement time $t$. For instance, here we report in Table \ref{tab:PL_explicit} the explicit expression of the $P^{(L)}(x,t)$ expansion for the lower orders $L=2,4,6,8$. The combinatorial structure mirrors the hierarchy of cumulant corrections in classical Edgeworth expansions, with the Touchard polynomials $T_p(t)$ encoding the statistics of the random number of jumps. 

\begin{table}[p]
\centering

\begin{tabular}{c l}
\hline
$L$ & Perturbative expansion $P^{(L)}(x,t)$ \\
\hline \\[-6pt]

$2$ &
$
P^{(2)}(x,t)
\underset{|x|\to\infty}{\sim}
f(x)\Big[
t
+ t^2\,\frac{d_2(x)}{x^2}
\Big]
$
\\[14pt]

$4$ &
$
P^{(4)}(x,t)
\underset{|x|\to\infty}{\sim}
f(x)\Big[
t
+ t^2\Big(
\frac{d_2(x)}{x^2}
+
\frac{d_4(x)}{x^4}\mathbb{E}[x^4]
\Big)
+ \frac{t^3+t^2+t}{2}\,
\frac{d_2(x)^2}{x^4}
\Big]
$
\\[16pt]

$6$ &
$
P^{(6)}(x,t)
\underset{|x|\to\infty}{\sim}
f(x)\Big[
t
+ t^2\Big(
\frac{d_2(x)}{x^2}
+
\frac{d_4(x)}{x^4}\mathbb{E}[x^4]
+
\frac{d_6(x)}{x^6}\mathbb{E}[x^6]
\Big)
+ \frac{t^3+t^2+t}{2}\Big(
\frac{d_2(x)^2}{x^4}
+
2\,\frac{d_2(x)d_4(x)}{x^6}\mathbb{E}[x^4]
\Big)
+ \frac{t^4+3t^3+t^2+t}{6}\,
\frac{d_2(x)^3}{x^6}
\Big]
$
\\[18pt]

$8$ &
$
P^{(8)}(x,t)
\underset{|x|\to\infty}{\sim}
f(x)\Big[
t
+ t^2\Big(
\frac{d_2(x)}{x^2}
+
\frac{d_4(x)}{x^4}\mathbb{E}[x^4]
+
\frac{d_6(x)}{x^6}\mathbb{E}[x^6]
+
\frac{d_8(x)}{x^8}\mathbb{E}[x^8]
\Big)
$
\\[6pt]
&
$
\quad\qquad\qquad\qquad\qquad
+ \frac{t^3+t^2+t}{2}\Big(
\frac{d_2(x)^2}{x^4}
+
2\,\frac{d_2(x)d_4(x)}{x^6}\mathbb{E}[x^4]
+
2\,\frac{d_2(x)d_6(x)}{x^8}\mathbb{E}[x^6]
+
\frac{d_4(x)^2}{x^8}\mathbb{E}[x^4]^2
\Big)
$
\\[6pt]
&
$
\quad\qquad\qquad\qquad\qquad
+ \frac{t^4+3t^3+t^2+t}{6}\Big(
\frac{d_2(x)^3}{x^6}
+
3\,\frac{d_2(x)^2 d_4(x)}{x^8}\mathbb{E}[x^4]
\Big)
$
\\[6pt]
&
$
\quad\qquad\qquad\qquad\qquad
+ \frac{t^5+6t^4+7t^3+t^2+t}{24}\,
\frac{d_2(x)^4}{x^8}
\Big]
$
\\[6pt]

\hline
\end{tabular}
\caption{Explicit perturbative expansion of the CTRW propagator $P(x,t)$ for exponential waiting times in the large $x$ limit for the perturbative orders $L=2,4,6,8$. The expressions are written entirely in terms of the perturbative coefficients $d_{l}(x)$, the moments $\mathbb{E}[x^{l}]$ of the jump distribution, and the Touchard polynomials of the Poisson process, where the time dependence is fully unrolled.}
\label{tab:PL_explicit}
\end{table}

\begin{figure}[ht]
\centering
\includegraphics[width=\textwidth]{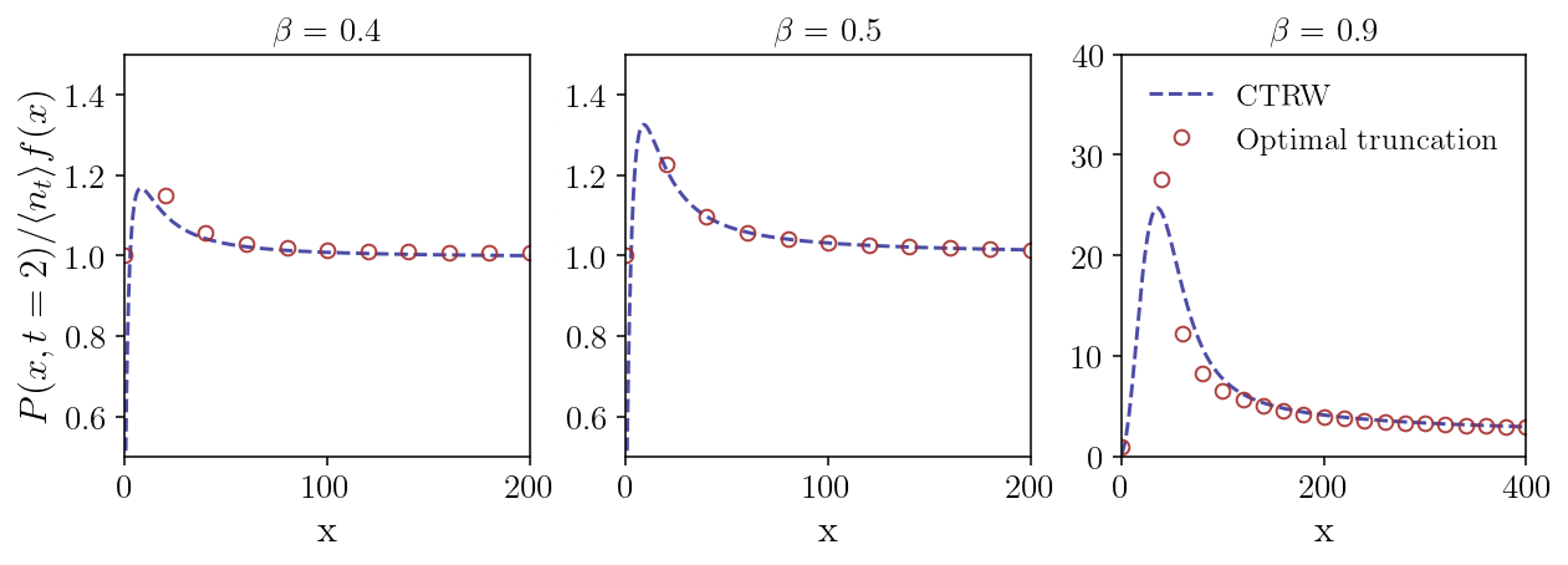}
\caption{Plot of the numerical simulation of a CTRW with exponential waiting times and sub-exponential stretched jump lengths, with unitary $\langle \tau \rangle$ and $\beta=0.4$ (left), $\beta=0.5$ (middle), $\beta=0.9$ (right), versus our optimal truncated perturbative expansion of the propagator $P(x,t)$, as prescribed by Eq. \eqref{P_x_t_poisson}. The blue dotted line represents the numerical simulation of the CTRW, while the red circles represent the optimal truncation method of Eq. \eqref{P_x_t_poisson}, simply substituting $G(x)$ with $G^{(L^*)}(x)$, following the procedure described in Subsection \ref{Subsection3C}.}
\label{Figure6}
\end{figure}
 
If we combine this perturbative development with the optimal truncation procedure described in Sec. \ref{Section3}, simply substituting $G(x)$ with $G^{(L^*)}(x)$ in Eq. \eqref{P_x_t_poisson}, we found an excellent agreement with numerical simulations, as one can see in Figure \ref{Figure6}. So the Poissonian statistics of the number of jumps is mapped, via our perturbative expansion, into a hierarchy of corrections to the BJP regime of the CTRW propagator.

\section{Conclusions}

In this work we have developed a perturbative framework that systematically extends the BJP to describe moderate deviations in sums of IID stretched–exponential random variables and in decoupled CTRWs. Starting from the geometric structure of the convolution integral, whose non-analytic cusps encode the $n$ distinct realizations of the big jump event, we constructed an expansion in powers of $1/x^2$ around the BJP configuration. The zeroth-order term naturally recovers the asymptotic BJP, whereas higher orders quantify the collective fluctuations of the remaining $n-1$ small jumps. This yields an explicit and controlled description of the large–$x$ behaviour of the conditional PDF $\phi(x|n)$, including its asymptotic structure, its domain of validity, and a practical optimal-truncation procedure. A central conceptual outcome is that the leading correction to the BJP already contains the scaling structure governing the crossover between Gaussian typical fluctuations and the condensation tail regime. Rather than emerging from matching arguments between different asymptotic theories, this scaling appears directly from the internal structure of the large $x$ perturbative expansion. From this term we derived a simplified yet accurate approximation of the anomalous rate function in the far-tail regime. While this result is consistent with large-deviation theory \cite{LDT_4, LDT_5}, the perturbative construction developed here does not rely on the large $n$ limit and remains valid for finite — and even small — numbers of summands. In this sense, our framework plays a role analogous to that of the Edgeworth expansion for the CLT: it provides a systematic and practically useful correction scheme around an asymptotic principle, independent of whether an exact large-deviation description is available.
The comparison with large-deviation approaches is therefore complementary. Recent work \cite{LDT_Perturbative_Naftali} developed a perturbative expansion in the opposite regime, starting from the CLT and expanding in large $n$. Together, these two perspectives describe the same crossover from opposite directions: a large $x$ expansion around condensation phase, and a large $n$ expansion around typical fluctuations. Our results show that much of the crossover structure can already be captured directly at finite $n$ through the perturbative correction of the BJP.
Beyond sums of IID variables, our perturbative expansion extends naturally to CTRWs through the subordination relation. This yields an analytical hierarchy of corrections to the BJP form of the propagator $P(x,t)$, expressed in terms of the moments $\langle n_t^p\rangle$ of the random number of jumps. For exponential waiting times, this structure becomes fully explicit and shows excellent agreement with numerical simulations when combined with optimal truncation. The method therefore provides a practical tool for predicting moderate-deviation properties of transport processes with stretched-exponential jumps. 
More broadly, the framework developed here isolates and quantifies the fluctuations around the big jump configuration itself. Because this mechanism underlies condensation phenomena in a wide class of stochastic systems, the perturbative structure may offer a natural starting point for extensions beyond the IID setting, including interacting systems where condensation competes with correlations \cite{BJP_Conclusions}. In this way, perturbative expansions around rare-event mechanisms may provide a general route toward controlled descriptions of fluctuation regimes that lie between typical behavior and extreme events.

\section{Acknowledgements}

A.B. and O.H. are grateful to Raffaella Burioni, Alessandro Vezzani, and Eli Barkai for insightful discussions, valuable comments, and encouragement throughout the development of this work. A.B. and O.H. also warmly thank the organizers of the Les Houches Summer School 2024 on Large Deviations and Applications for creating an inspiring scientific environment. It was during this school that the authors first met and began their scientific collaboration, which ultimately led to the present work.

\appendix
\counterwithin{figure}{section}

\section{Symmetry of the cusps \label{AppendixA}}

In this appendix we show that any non-central cusp gives the same leading contribution as the central one.  Without loss of generality, we consider the cusp located at $\vec y=(1,0,\ldots,0)$. To isolate its local contribution, we restrict the integration domain so that $y_1$ lies in the region containing this cusp, while all other coordinates remain in the region associated with the central one. 
For definiteness we split the domains at a finite value (e.g. $1/2$), which does not affect the leading asymptotics. Near this cusp the function $g(\vec y)$ reads
\[
g(\vec y)\approx \sum_{i=2}^{n-1}|y_i|^\beta + 1 + |1-y_1|^\beta .
\]
Using Eqs.~(\ref{convolution},\ref{g_y_n}), the corresponding contribution is
\begin{align}
\mathrm{cusp}
\approx&
|x|^{n-1}\widetilde N^{\,n}
\int_{-\infty}^{1/2}dy_{n-1}\cdots\int_{-\infty}^{1/2}dy_{2}
\exp\!\left[-\alpha^\beta |x|^\beta \sum_{i=2}^{n-1}|y_i|^\beta\right]
\nonumber\\
&\times
\int_{1/2}^{\infty}dy_{1}
\exp\!\left[-\alpha^\beta |x|^\beta\left(1+|1-y_1|^\beta\right)\right]
\nonumber\\[6pt]
=&
|x|^{n-1}\widetilde N^{\,n}
e^{-\alpha^\beta|x|^\beta}
\left[
\int_{-\infty}^{1/2}dy\,
e^{-\alpha^\beta|x|^\beta |y|^\beta}
\right]^{n-2}
\int_{1/2}^{\infty}dy_1\,
e^{-\alpha^\beta|x|^\beta |1-y_1|^\beta}.
\end{align}
In the last integral we perform the change of variables $y'_1=1-y_1$, which gives
\[
\int_{1/2}^{\infty}dy_1\,e^{-\alpha^\beta|x|^\beta |1-y_1|^\beta}
=
\int_{-\infty}^{1/2}dy'_1\,e^{-\alpha^\beta|x|^\beta |y'_1|^\beta}.
\]
Hence all integrals coincide, and defining $
\mathcal A = \int_{-\infty}^{1/2}dy\,
e^{-\alpha^\beta|x|^\beta |y|^\beta},
$ we obtain
\begin{equation}
  \mathrm{cusp}
=
\left(|x|\widetilde N\mathcal A\right)^{n-1}f(x).
\end{equation}
For large $x$, the integral $\mathcal A$ is dominated by a narrow region around the origin and coincides, to leading order, with the normalization integral of the jump distribution. Therefore$|x|\widetilde N\mathcal A \simeq 1$,
and the contribution of this cusp reduces to
$\mathrm{cusp}\sim f(x)$. Thus any non-central cusp yields the same leading contribution as the central one. Since there are $n$ equivalent cusps, the total large $x$ contribution is $n f(x)$, consistently with the BJP.

\section{Calculation of $d_l(x)$ \label{AppendixB}}

In this appendix we show the complete derivation of the coefficients $d_l(x)$ in the case of $n=2$ starting from equation \eqref{phi_x_2_series}. The result we obtain for $d_l(x)$ at each order extends to the generic $n$ case. Let us focus on the perturbative exponential in \eqref{phi_x_2_exponential_expansion}, using a different index $k$ for the sake of clarity, and write it as a series of powers:

\begin{align}
{\rm exp}\left[-\alpha^\beta|x|^\beta\sum_{k=1}^{\infty}(-1)^k\binom{\beta}{k}y^k\right] =& \sum_{j=0}^{\infty} \frac{\left(-\alpha^\beta|x|^\beta\sum_{k=1}^{\infty}(-1)^k\binom{\beta}{k}y^k\right)^{j}}{j!} \\ \nonumber
=& \sum_{j=0}^{\infty} \frac{(-1)^j}{j!}\alpha^{j\beta}|x|^{j\beta}\left( \sum_{k=1}^{\infty} (-1)^k \binom{\beta}{k} y^k \right)^j 
\end{align}

\noindent We truncate our series expansion in the limit of $x$ big to a generic finite order $L$. Thus, the exponential becomes:

\begin{align}
{\rm exp}\left[-\alpha^\beta|x|^\beta\sum_{k=1}^{\infty}(-1)^k\binom{\beta}{k}y^k\right] \approx& \sum_{j=0}^{L} \frac{(-1)^j}{j!}\alpha^{j\beta}|x|^{j\beta}\left( \sum_{k=1}^{L} (-1)^k \binom{\beta}{k} y^k \right)^j \\ \nonumber
\approx& \sum_{j=0}^L \frac{(-1)^j}{j!}\alpha^{j\beta}|x|^{j\beta} \left( -\binom{\beta}{1}y + \binom{\beta}{2}y^2 - .... + (-1)^L\binom{\beta}{L}y^L  \right)^j
\end{align}

\noindent Focusing on the sum of $L$ terms raised to power $j$, we can use Newton's Multinomial theorem \cite{Multinomial_thm} to unroll the sum:

\begin{equation}
    \left( -\binom{\beta}{1}y +  .... + (-1)^L\binom{\beta}{L}y^L  \right)^j = \sum_{\substack{\text{$j_1, ..., j_L$} \\ \text{$j_1 + ... + j_L = j$}}} \frac{j!}{j_1! ... j_L!}(-1)^{j_1 +... + Lj_L}\binom{\beta}{1}^{j_1} y^{j_1} ... \binom{\beta}{L}^{j_L}y^{j_L L} 
\end{equation}
So, the exponential becomes a truncated set of powers, and since the truncation is at $L$-th order, we need to constrain the sum on the indices $j_1, ..., j_L$, imposing that the product $y^{j_1} y^{2j_2} ...y^{Lj_L}$ satisfies $j_1 + 2j_2 + ... + Lj_L \leq L$. As a consequence we have:

\begin{align}
{\rm exp}\left[-\alpha^\beta|x|^\beta\sum_{k=1}^{\infty}(-1)^k\binom{\beta}{k}y^k\right] \approx&\sum_{j=0}^{L} \sum_{\substack{\text{$j_1+ ... + j_L=j$} \\ \text{$j_1 + ... + Lj_L \leq L$}}} \frac{(-1)^{j+j_1 + ... + Lj_L}}{j_1! ... j_L!} \alpha^{j\beta}|x|^{j\beta}\binom{\beta}{1}^{j_!} ... \binom{\beta}{L}^{j_L} y^{j_1 + ... + Lj_L}
\end{align}

Now, returning to Eq. \eqref{phi_x_2_exponential_expansion} and inserting our power series, we see the emergence of the coefficients $d_l(x)$:
\begin{align}
\phi(x|2) &\approx  2\widetilde{N}^2 |x| \int_{-\infty}^{\infty} {\rm e}^{\alpha^\beta|x|^\beta (-1-|y|^\beta)} \times  \\ \nonumber
& \ \  \left(\sum_{j=0}^{L} \sum_{\substack{\text{$j_1+ ... + j_L=j$} \\ \text{$j_1 + ... + Lj_L \leq L$}}} \frac{(-1)^{j+j_1 + ... + Lj_L}}{j_1! ... j_L!} \alpha^{j\beta}|x|^{j\beta}\binom{\beta}{1}^{j_!} ... \binom{\beta}{L}^{j_L} y^{j_1 + ... + Lj_L}\right)dy \\ \nonumber
&\approx 2\widetilde{N}^2 |x| e^{-\alpha^{\beta}|x|^{\beta}} \int_{-\infty}^{+\infty} e^{-\alpha^{\beta}|x|^{\beta}|y|^{\beta}}\left( 1 + \text{d}_1(x) y + \text{d}_2(x) y^2 + ... + \text{d}_L(x)y^L \right) dy
\end{align}
Where here $d_l(x)$ denote all the prefactors of the sum of powers of $y$ having a combination of positive indices $(j_1, ..., j_L)$ such that the constraints $j_1 + ... + j_L = j$ and $j_1 +... + Lj_L=l$, for $j=1,2,...,l$ and $l \leq L$ are satisfied. Recalling the change of variable $y=\chi/x$, we rewrite the integral in terms of $\chi$:

\begin{align}
\phi(x|2) &\approx 2\widetilde{N}^2 |x| e^{-\alpha^{\beta}|x|^{\beta}} \int_{-\infty}^{+\infty} e^{-\alpha^{\beta}|\chi|^{\beta}}\left( 1 + \frac{\text{d}_1(x)}{x}\chi + \frac{\text{d}_2(x)}{x^2} \chi^{\ 2} + ... + \frac{\text{d}_L(x)}{x^L}\chi^{\ L} \right) \frac{d\chi}{|x|} \\  \nonumber
&\approx 2\widetilde{N}^2 e^{-\alpha^{\beta}|x|^{\beta}} \int_{-\infty}^{+\infty} \frac{f(\chi)}{\widetilde{N}} \left( 1 + \frac{\text{d}_1(x)}{x}\chi + \frac{\text{d}_2(x)}{x^2} \chi^{\ 2} + ... + \frac{\text{d}_L(x)}{x^L}\chi^{\ L} \right) d\chi \\ \nonumber 
&\approx 2\widetilde{N}e^{-\alpha^{\beta}|x|^{\beta}}\int_{-\infty}^{+\infty} f(\chi) \left( 1 + \frac{\text{d}_1(x)}{x}\chi + \frac{\text{d}_2(x)}{x^2} \chi^{\ 2} + ... + \frac{\text{d}_L(x)}{x^L}\chi^{\ L} \right) d\chi \\ \nonumber
&\approx 2 f(x)\bigg( \int_{-\infty}^{+\infty} f(\chi) d\chi + \frac{ \text{d}_1(x)}{x} \int_{-\infty}^{+\infty} \chi f(\chi) d\chi  + \frac{\text{d}_2(x)}{x^2} \int_{-\infty}^{+\infty} \chi^{\ 2}f(\chi) d\chi+ \nonumber \\ &\ \ \ ... + \frac{\text{d}_L(x)}{x^L} \int_{-\infty}^{+\infty} \chi^{\ L}f(\chi) \bigg) d\chi \\ \nonumber
&\approx 2 f(x) \left(1 + \frac{\text{d}_1(x)}{x}\mathbb{E}[x] + \frac{\text{d}_2(x)}{x^2}\mathbb{E}[x^2] + ... + \frac{\text{d}_L(x)}{x^L}\mathbb{E}[x^L]\right)
\end{align}

Now the final result for $\text{d}_l(x)$, where $l$ can be at this stage either even either odd, is given by:

\begin{equation}
\label{d_l_x_general}
\text{d}_l(x) = \sum_{j=1}^l \sum_{\substack{\text{$j_1 + j_2 + ... +j_n = j$} \\ \text{$j_1 + 2j_2 + ... + nj_n = l$}}} \frac{(-1)^{j+l}}{j_1! j_2! ... j_l!} \alpha^{j\beta} |x|^{j\beta} \binom{\beta}{1}^{j_1} \binom{\beta}{2}^{j_2} ... \binom{\beta}{l}^{j_l}
\end{equation}

We now use the even symmetry of the stretched exponential function $f(x)$, i.e., we use the fact that all its moments of odd order, and let us recall their value:

\begin{equation}
\mathbb{E}[x^l] = 
\begin{cases}
0 & \text{if $l$ odd} \\
\frac{\Gamma\left(1/\beta\right)^{l/2 - 1}}{\Gamma(3/\beta)^{l/2}}\Gamma\left( \frac{l+1}{2} \right) & \text{if $l$ even}
\end{cases}
\end{equation}

Consequently, we proved the final result of formula \eqref{phi_x_2_series}, and the sum along $l$ due to the symmetry of the moments is only on even terms. 

\section{Calculation of mixed terms \label{AppendixC}}

In this appendix we show the calculations to obtain the full approximation of $\phi(x|n)$ taking into account also the mixed terms. We start from Eq. \eqref{multinomial_dev}, and explicate the various pre-factors obtained by applying Newton's Multinomial theorem \cite{Multinomial_thm}:

\begin{equation}
\left( \sum_{i=1}^{n-1} y_i \right)^l = \sum_{i=1}^{n-1} y_i^l + \sum_{\substack{\text{$l_1, l_2$} \\ \text{$l_1+l_2=l$}}} \frac{l!}{l_1! l_2!} \sum_{\substack{\text{$i_1, i_2$} \\ \text{$i_1 \neq i_2$}}}^{n-1} y_{i_1}^{l_1} y_{i_2}^{l_2} + ... + \sum_{\substack{\text{$l_1, ..., l_m$} \\ \text{$l_1+... +l_m=l$}}} \frac{l!}{l_1! ... l_m!} \sum_{\substack{\text{$i_1, ... i_m$} \\ \text{$i_1 \neq ... \neq i_m$}}}^{n-1} y_{i_1}^{l_1} ... y_{i_m}^{l_m}
\end{equation}
Where $m=\min(l,n-1)$. The key point is to identify that this sum can be splitted in two different parts, i.e. coherent part and mixed part, as explained in Subsection \ref{Subsection3B}. We can write it as:
\begin{align}
    & \left( \sum_{i=1}^{n-1} y_i \right)^l = \text{Coh}^l + \text{Mix}
    ^l \\ \nonumber
    & \ \ \text{Coh}^l =  \sum_{i=1}^{n-1} y_i^l \\ \nonumber
    & \ \ \text{Mix}^l = \sum_{\substack{\text{$l_1, l_2$} \\ \text{$l_1+l_2=l$}}} \frac{l!}{l_1! l_2!} \sum_{\substack{\text{$i_1, i_2$} \\ \text{$i_1 \neq i_2$}}}^{n-1} y_{i_1}^{l_1} y_{i_2}^{l_2} + ... + \sum_{\substack{\text{$l_1, ..., l_m$} \\ \text{$l_1+... +l_m=l$}}} \frac{l!}{l_1! ... l_m!} \sum_{\substack{\text{$i_1, ... i_m$} \\ \text{$i_1 \neq ... \neq i_m$}}}^{n-1} y_{i_1}^{l_1} ... y_{i_m}^{l_m}
\end{align}
So, the approximated function $g(\vec{y})$ around $\vec{y}=\vec{0}$ becomes in its completeness:
\begin{eqnarray}
   g(\vec{y})\approx \sum_{i=1}^{n-1}|y_i|^\beta+1+\sum_{l=1}^{\infty}(-1)^l\binom{\beta}{l}\text{Coh}^l +\sum_{l=1}^{\infty}(-1)^l\binom{\beta}{l}\text{Mix}^l
\end{eqnarray}
So, the approximation of $\phi(x|n)$ in its full expression becomes:
\begin{align}
    \phi(x|n) \approx n f(x)|x\widetilde{N}|^{n-1} &\int\cdots\int dy_{1}..dy_{n-1}\exp\left[-\alpha^\beta|x|^\beta \sum_{i=1}^{n-1}|y_i|^\beta\right] \\ \nonumber
    &\times \exp\left[-\alpha^\beta|x|^\beta \sum_{l=1}^{\infty}(-1)^l\binom{\beta}{l}(\text{Coh}^l+\text{Mix}^l)\right]
\end{align}
As already proved in Appendix \ref{AppendixB},  calling $\gamma_l =\alpha^\beta|x|^\beta(-1)^l \binom{\beta}{l}$ for simplicity, considering the coherent part $\text{Coh}^l$ we obtain:
\begin{align}
    &\int\cdots\int dy_{1}..dy_{n-1}\exp\left[-\alpha^\beta|x|^\beta \sum_{i=1}^{n-1}|y_i|^\beta\right]\exp\left[- \sum_{l=1}^{\infty}\gamma_l\text{Coh}^l\right] \\ \nonumber
    =&\int\cdots\int dy_{1}..dy_{n-1}\exp\left[-\alpha^\beta|x|^\beta \sum_{i=1}^{n-1}|y_i|^\beta\right]\exp\left[- \sum_{l=1}^{\infty}\gamma_l\left( \sum_{i=1}^{n-1}y_i\right)^l\right]
    \\ \nonumber
    =&\int\cdots\int dy_{1}..dy_{n-1}\prod_{i=1}^{n-1}\exp\left[-\alpha^\beta|x|^\beta |y_i|^\beta\right]\exp\left[- \sum_{l=1}^{\infty}\gamma_l y_i^l\right]\\ \nonumber
    =&\left(1 + \sum_{\text{$l$ even}}^{\infty} \frac{d_l(x) }{x^l} \mathbb{E}[x^l] \right)^{n-1}\\ \nonumber
    =&   \ \mathrm{G}(x)^{n-1}
\end{align}
Dividing the coherent and the mixed contributions, we have:
\begin{equation}
    \int\cdots\int dy_{1}..dy_{n-1}\exp\left[-\alpha^\beta|x|^\beta \sum_{i=1}^{n-1}|y_i|^\beta\right]\exp\left[- \sum_{l=1}^{\infty}\gamma_l\text{Coh}^l\right]\exp\left[- \sum_{l=1}^{\infty}\gamma_l\text{Mix}^l\right]
\end{equation}
Now, in first approximation we can develop the exponential of the mixed terms as $\exp\left[- \sum_{l=1}^{\infty}\gamma_k\text{Mix}^k\right] \sim 1 - \sum_{k=1}^{\infty}\gamma_k\text{Mix}^k$, since $x \rightarrow \infty$ (where we change the index from $l$ to $k$), and so $\text{Mix}^k(x) \rightarrow 0$ for every order of $k$. Returning into the initial variables, so $y_i = \chi_i / x$, we have that both coherent part and mixed part gain a prefactor in $x$, in particular $\text{Coh}^l \rightarrow \text{Coh}^l/x^l$ and $\text{Mix}^k \rightarrow \text{Mix}^k/x^k$. The $n-1$ convolution then becomes:
\begin{align}
    &  \ \ \ \ \   \int\cdots\int d\chi_{1}..d\chi_{n-1}\exp\left[-\alpha^\beta \sum_{i=1}^{n-1}|\chi_i|^\beta\right]\exp\left[- \sum_{l=1}^{\infty}\frac{\gamma_l}{x^l}\text{Coh}^l\right]\exp\left[- \sum_{k=1}^{\infty}\frac{\gamma_k}{x^k}\text{Mix}^k\right] \\ \nonumber
    &\sim\int\cdots\int d\chi_{1}..d\chi_{n-1}\prod_{i=1}^{n-1}\exp\left[-\alpha^\beta |\chi_i|^\beta\right]\exp\left[- \sum_{l=1}^{\infty}\frac{\gamma_l}{x^l}\chi_i^l\right]\left(1 - \sum_{k=1}^{\infty}\frac{\gamma_k}{x^k}\text{Mix}^k\right)\\ \nonumber
    &\sim\int\cdots\int d\chi_{1}..d\chi_{n-1}\prod_{i=1}^{n-1}f(\chi_i)\left(1 + \sum_{l  \ \text{even}}^{\infty} \frac{d_l(x)}{x^l}\chi_i^l\right)\left(1 - \sum_{k=1}^{\infty}\frac{\gamma_k}{x^k}\text{Mix}^k\right)
\end{align}
Introducing the notation:
\begin{equation}
    g^{(l)}(\chi) = 1 + \sum_{l  \ \text{even}}^{\infty} \frac{d_l(x)}{x^l}\chi^l 
\end{equation}
It yields:
\begin{align}
& \ \ \ \ \int\cdots\int d\chi_{1}..d\chi_{n-1}\prod_{i=1}^{n-1}f(\chi_i)g^{(l)}(\chi_i)\left(1 - \sum_{k=1}^{\infty}\frac{\gamma_k}{x^k}\text{Mix}^k\right) \\ \nonumber
&\sim G(x)^{n-1} - \int\cdots\int d\chi_{1}..d\chi_{n-1}\prod_{i=1}^{n-1}f(\chi_i)g^{(l)}(\chi_i) \left(\sum_{k=1}^{\infty}\frac{\gamma_k}{x^k}\text{Mix}^k\right)
\end{align}
So, in terms of the new notation the result for $\phi(x|n)$ becomes:

\begin{equation}
    \phi(x|n) \sim nf(x) \left[ G(x)^{n-1} - \int\cdots\int d\chi_{1}..d\chi_{n-1}\prod_{i=1}^{n-1}f(\chi_i)g^{(l)}(\chi_i) \left(\sum_{k=1}^{\infty}\frac{\gamma_k}{x^k}\text{Mix}^k\right) \right]
\end{equation}
Where the first term is the approximated result without mixed terms, i.e. without "interactions" of other jumps, that coincides with the result \eqref{phi_x_n_pert} obtained in Section \ref{Subsection3B}, and the second term is the first order correction due to the mixed terms. The focus here is on the second integral. Using their explicit definition, it becomes:
\begin{align}
\label{ugly_integral_mix}
    \int ...\int & \prod_{i=1}^{n-1} d\chi_i f(\chi_i) g^{(l)}(\chi_i)\bigg[\sum_{k=1}^{\infty} \frac{\gamma_k}{x^k} \bigg(\sum_{\substack{\text{$k_1, k_2$} \\ \text{$k_1+k_2=k$}}} \frac{k!}{k_1! k_2!} \sum_{\substack{\text{$i_1, i_2$} \\ \text{$i_1 \neq i_2$}}}^{n-1} \chi_{i_1}^{k_1} \chi_{i_2}^{k_2} +\\ \nonumber
    & ... + \sum_{\substack{\text{$k_1, ..., k_m$} \\ \text{$k_1+... +k_m=k$}}} \frac{k!}{k_1! ... k_m!} \sum_{\substack{\text{$i_1, ... i_m$} \\ \text{$i_1 \neq ... \neq i_m$}}}^{n-1} \chi_{i_1}^{k_1} ... \chi_{i_m}^{k_m}\bigg)\bigg]
\end{align}
The integrals in $\chi_i$, ignoring the numerical prefactors, contain the interactions of binary terms, that modify only two variables, of ternary terms, that act only on three variables, etc. Taking care only of integrations, we see for example the case of binary terms:

\begin{align}
    &\int ... \int  \prod_{i=1}^{n-1} d\chi_i f(\chi_i) g^{(l)}(\chi_i) \chi_{i_1}^{k_1}\chi_{i_2}^{k_2} \\ \nonumber
    &= \int...\int\prod_{\substack{\text{$i=1$} \\ \text{$i\neq i_1, i_2$}}}^{n-3} d\chi_i f(\chi_i) g^{(l)}(\chi_i) \int d\chi_{i_1} f(\chi_{i_1}) g^{(l)}(\chi_{i_1}) \chi_{i_1}^{k_1} \int d\chi_{i_2} f(\chi_{i_2}) g^{(l)}(\chi_{i_2}) \chi_{i_2}^{k_2} \\ \nonumber
    &=x^{k_1+k_2}G(x)^{n-3} \int d\chi_{i_1} f(\chi_{i_1}) g_{k_1}^{(l)}(\chi_{i_1}) \int d\chi_{i_2} f(\chi_{i_2}) g_{k_2}^{(l)}(\chi_{i_2}) \\ \nonumber
    &\equiv x^{k_1+k_2}G(x)^{n-3}G_{k_1}(x)G_{k_2}(x)
\end{align}
Where here the notation $g_k^{(l)}(\chi)$ and $G_k(x)$ means:

\begin{equation}
     g_k^{(l)}(\chi)=\frac{d_l(x)}{x^{l+k}}\chi^{l+k},\qquad G_k(x)= \left(\frac{\mathbb{E}[x^k]}{x^k} + \sum_{\substack{\text{$l>0$} \\\text{$l+k$ even}}}\frac{d_l(x)}{x^{l+k}}\mathbb{E}[x^{l+k}]\right)
\end{equation}
Considering the binary case as a prototypical example, considering the factor $x^{k_1+k_2}$ we have that the integral is multiplied by $\gamma_k/x^k$, and $k_1, k_2$ are two positive indexes such that $k_1+k_2=k$. So the prefactor in $x$ will disappear in the final expression of $\phi(x|n)$. As a consequence it's possible to generalize this approach to all other successive interactions:

\begin{align}
   &\frac{1}{x^k}\int ... \int  \prod_{i=1}^{n-1} d\chi_i f(\chi_i) g^{(l)}(\chi_i) \chi_{i_1}^{k_1}\chi_{i_2}^{k_2} = G(x)^{n-3}G_{k_1}(x)G_{k_2}(x) \\ \nonumber
   &\frac{1}{x^k}\int ... \int  \prod_{i=1}^{n-1} d\chi_i f(\chi_i) g^{(l)}(\chi_i) \chi_{i_1}^{k_1}\chi_{i_2}^{k_2}\chi_{i_3}^{k_3} = G(x)^{n-4}G_{k_1}(x)G_{k_2}(x)G_{k_3}(x) \\ \nonumber
   &...\\ \nonumber
   &\frac{1}{x^k}\int ... \int  \prod_{i=1}^{n-1} d\chi_i f(\chi_i) g^{(l)}(\chi_i) \chi_{i_1}^{k_1}\chi_{i_2}^{k_2}...\chi_{i_m}^{k_m} = G(x)^{n-m-1}G_{k_1}(x)G_{k_2}(x)....G_{k_m}(x)
\end{align}
Now, returning at Eq. \eqref{ugly_integral_mix}, it becomes:
\begin{align}
&\sum_{k=1}^{\infty}\gamma_k\bigg(\sum_{\substack{\text{$k_1, k_2$} \\ \text{$k_1+k_2=k$}}} \frac{k!}{k_1!k_2!}\sum_{\substack{\text{$i_1, i_2$} \\ \text{$i_1 \neq i_2$}}}^{n-1} \frac{1}{x^k}\int...\int \prod_{i=1}^{n-1} d\chi_i f(\chi_i) g^{(l)}(\chi_i) \chi_{i_1}^{k_1}\chi_{i_2}^{k_2} + ... \\ \nonumber
&+ \sum_{\substack{\text{$k_1,..,k_m$} \\ \text{$k_1+...+k_m=k$}}}\frac{k!}{k_1!...k_m!}\sum_{\substack{\text{$i_1,..,i_m$} \\ \text{$i_1 \neq ...\neq i_m$}}}^{n-1}\frac{1}{x^k}\int ... \int  \prod_{i=1}^{n-1} d\chi_i f(\chi_i) g^{(l)}(\chi_i) \chi_{i_1}^{k_1}\chi_{i_2}^{k_2}...\chi_{i_m}^{k_m}\bigg)\\ \nonumber
=& \sum_{k=1}^{\infty} \gamma_k \bigg( \sum_{\substack{\text{$k_1, k_2$} \\ \text{$k_1+k_2=k$}}} \frac{k!}{k_1!k_2!}\sum_{\substack{\text{$i_1, i_2$} \\ \text{$i_1 \neq i_2$}}}^{n-1} G(x)^{n-3}G_{k_1}(x)G_{k_2}(x) + ... \\ \nonumber
\ &+\sum_{\substack{\text{$k_1,..,k_m$} \\ \text{$k_1+...+k_m=k$}}}\frac{k!}{k_1!...k_m!}\sum_{\substack{\text{$i_1,..,i_m$} \\ \text{$i_1 \neq ...\neq i_m$}}}^{n-1}G(x)^{n-m-1}G_{k_1}(x)G_{k_2}(x)....G_{k_m}(x) \bigg)
\end{align}
Now, to lighten this cumbersome expression we can remove the sums over the indices $i_1, i_2... i_m$ since ours RVs $\chi_1, ...\chi_m$ are IID. We need then to address a combinatorial question: how many possible combinations there are to realize for instance a binary term $\chi_{i_1}^{k_1}\chi_{i_2}^{k_2}$ with the constraints $k_1+k_2=k$ and $i_1 \neq i_2$? It's possible to find a combinatorial rule for each successive interaction, which is:
\begin{align}
    \text{\# binary terms} \ \ & \ \chi_{i_1}^{k_1}\chi_{i_2}^{k_2} \ \ \ \ \ \binom{n-1}{2}(k-1) \\ \nonumber
    \text{\# ternary terms} \ \ & \ \chi_{i_1}^{k_1}\chi_{i_2}^{k_2}\chi_{i_3}^{k_3} \ \ \ \ \ \binom{n-1}{3}\binom{k-1}{2} \\ \nonumber
    ...\\ \nonumber
    \text{\# $m$-th order terms} \ \ & \ \chi_{i_1}^{k_1}\chi_{i_2}^{k_2}...\chi_{i_m}^{k_m} \ \ \ \ \ \binom{n-1}{m}\binom{k-1}{m-1}
\end{align}
So we can write explicitly the full expression for $\phi(x|n)$, taking into account the effects of the mixed terms:
\begin{align}
\label{mixed_terms}
\phi(x|n) \sim nf(x) \bigg[&G(x)^{n-1} - \\ \nonumber
&\alpha^{\beta}|x|^{\beta}\sum_{k=1}^{\infty} (-1)^k \binom{\beta}{k}\bigg(\binom{n-1}{2}(k-1)\sum_{\substack{\text{$k_1, k_2>0$} \\ \text{$k_1+k_2=k$}}} \frac{k!}{k_1!k_2!} G(x)^{n-3}G_{k_1}(x)G_{k_2}(x)+...\\ \nonumber  
&+\binom{n-1}{m}\binom{k-1}{m-1}\sum_{\substack{\text{$k_1,..,k_m$} \\ \text{$k_1+...+k_m=k$}}}\frac{k!}{k_1!...k_m!}G(x)^{n-m-1}G_{k_1}(x)G_{k_2}(x)....G_{k_m}(x) \bigg) \bigg]  
\end{align}

In conclusion, here we show that all mixed terms can be suppressed at every perturnative order $l$ with a simple scaling argument. The full expression \eqref{mixed_terms} involves contributions for mixed terms of the form:
\begin{equation}
\alpha^\beta |x|^\beta
\sum_{k=1}^{\infty} (-1)^k \binom{\beta}{k}
\sum_{m=2}^{k}
\binom{n-1}{m}\binom{k-1}{m-1}
\sum_{\substack{k_1,\dots,k_m \ge 1 \\ k_1+\dots+k_m = k}}
\frac{k!}{k_1!\cdots k_m!}\,
G(x)^{\,n-m-1}\,
\prod_{i=1}^{m} G_{k_i}(x),    
\end{equation}
where mixed terms correspond to $m\ge2$, where $m=\min(l,n-1)$. To assess their relevance, we recall the large $x$ leading behavior of the two building blocks:
\begin{equation}
G(x) \sim O\!\left(|x|^{-2(1-\beta)}\right),
\qquad
G_k(x)
\sim O\!\left(|x|^{-k-2(1-\beta)}\right),
\quad k\ge1,  
\end{equation}
Finally, coherent terms at order $l$ scale as
$|x|^{-l(2-2\beta)}$
whereas mixed terms at the same order scale as $|x|^{-k-2l(1-\beta)}$, and since $0<\beta<1$, we have that $l(2-2\beta) < k+l(2-2\beta)$ for every positive $l$ and $k\leq l$, showing that mixed terms are always suppressed by an extra factor of order $|x|^{-k}$. So, at every perturbative order $l$, mixed contributions are strictly sub-leading compared to the coherent sector encoded in $G(x)^{\,n-1}$.  Thus, the dominant asymptotic behavior of the conditional PDF is fully captured by \eqref{phi_x_n_pert},
while mixed terms describe correlated small-jump fluctuations and enter only as finite-size corrections.

\bibliography{BJ_corrections_biblio}

\end{document}